\documentclass[pra,twocolumn,
                                groupedaddress,
                                showpacs,
                                ]{revtex4}
\usepackage{graphicx}
\usepackage{amsmath}
\usepackage{amsfonts}
\usepackage{amssymb}
\usepackage{hyperref}

\newcommand{\beq}{\begin{equation}}
\newcommand{\eeq}{\end{equation}}

\usepackage{placeins} %%% 02/12/14
\usepackage{bm}         % % % 05/12/14

\begin{document}

\title{Mean-field theory of atomic self-organization in optical cavities}

\author{Simon B. J\"ager} 
\affiliation{Theoretische Physik, Universit\"at des Saarlandes, D-66123 Saarbr\"ucken, Germany} 

\author{Stefan Sch\"utz} 
\affiliation{Theoretische Physik, Universit\"at des Saarlandes, D-66123 Saarbr\"ucken, Germany}

\author{Giovanna Morigi} 
\affiliation{Theoretische Physik, Universit\"at des Saarlandes, D-66123 Saarbr\"ucken, Germany} 

\date{\today}

\begin{abstract}
Photons mediate long-range optomechanical forces between atoms in high finesse resonators, which can induce the formation of ordered spatial patterns. When a transverse laser drives the atoms, the system undergoes a second order phase transition, that separates a uniform spatial density from a Bragg grating maximizing scattering into the cavity and is controlled by the laser intensity. Starting from a Fokker-Planck equation describing the  semiclassical dynamics of the $N$-atom distribution function, we systematically develop a mean-field model and analyse its predictions for the equilibrium and out-of-equilibrium dynamics. The validity of the mean-field model is tested by comparison with the numerical simulations of the $N$-body Fokker-Planck equation and by means of a BBGKY hierarchy. The mean-field theory predictions well reproduce several results of the $N$-body Fokker-Planck equation for sufficiently short times, and are in good agreement with existing theoretical approaches based on field-theoretical models. Mean-field, on the other hand, predicts thermalization time scales which are at least one order of magnitude shorter than the ones predicted by the $N$-body dynamics. We attribute this discrepancy to the fact that the mean-field ansatz discards the effects of the long-range incoherent forces due to cavity losses.  % and exhibit a linear scaling with $N$, while when all correlations are accounted for the scaling of the relaxation time is instead superlinear. 
\end{abstract}

% 37.30.+i 	Atoms, molecules, and ions in cavities 
% 42.65-Sf 	Dynamics of nonlinear optical systems; optical instabilities, optical chaos and complexity, and optical spatio-temporal dynamics 
%05.65.+b	Self-organized systems 
%05.70.Ln 	 Nonequilibrium and irreversible thermodynamics 

\maketitle

\section{Introduction}

Optically-dense atomic ensembles offer a formidable framework to study collective effects induced by atom-photon interactions \cite{Domokos:2013,Oppo,Chang}. Correlations are established by multiple photon scattering \cite{ODell,Rempe}, which can give rise to phenomena such as synchronization \cite{Kuramoto,Holland}, optomechanical bistability \cite{Bistability,Ritter:2009}, and spontaneous spatial ordering \cite{Oppo,Chang,Domokos:2002,Black:2003,Baumann:2010}. Envisaged applications for these systems range from sensors \cite{Stamper-Kurn}, to quantum-enhanced metrology \cite{Reichel} and quantum simulators \cite{Baumann:2010,Nagy:2010}. %Moreover, the dynamics observed in these systems are expected to shed light into the dynamics of long-range interacting systems \cite{}. 

\begin{figure}
\includegraphics[width=1\linewidth]{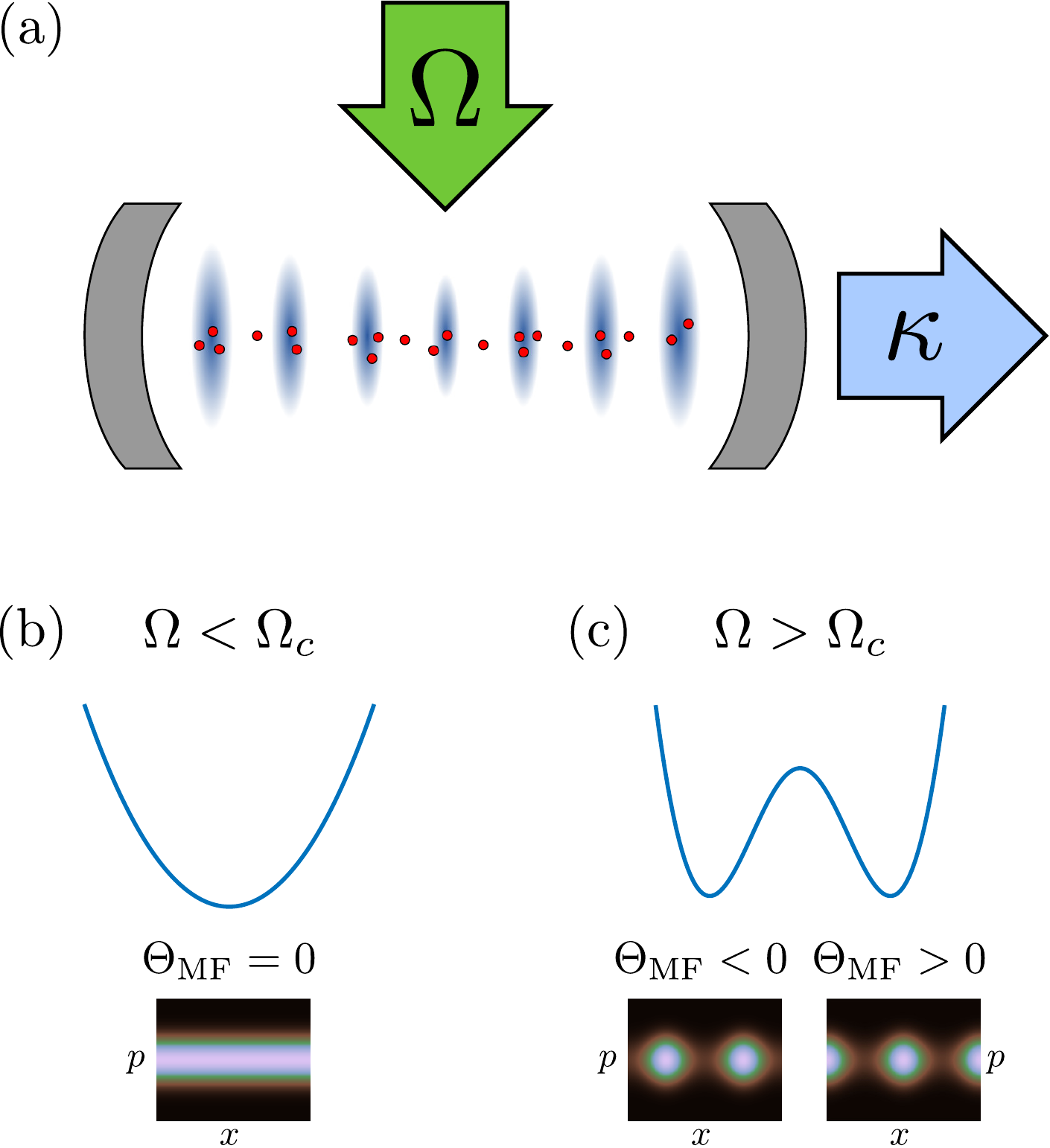}
\caption{(Color online) (a) Atoms in a standing-wave cavity and driven by a transverse laser can spontaneously form ordered patterns when the amplitude of the laser coupling $\Omega$, exceeds a threshold value $\Omega_c$, which depends on the rate of photon losses, here due to cavity decay at rate $\kappa$. In this regime the system undergoes a second-order phase transition which is characterised by the parameter $\Theta$, indicating spatial ordering of the atoms in Bragg gratings and defined in Eq. \eqref{Theta:N}. Its expectation value in the mean-field description is denoted as $\Theta_{\rm MF}$, see 
subplots (b) and (c), which display the thermodynamic potential below and above threshold. The lower panels are the single-particle density distribution $f_1(x,p)$ in phase space. In (b) the atomic density is uniform, in (c) it is localized at the even or odd sites of the cavity standing wave, ($\cos(kx)=1$ or $-1$, respectively). In this work we derive and discuss a mean-field theory for the dynamics of $f_1(x,p)$. }
\label{fig:1}
\label{cavity}
\end{figure}

Single-mode cavities, furthermore, mediate strong long-range interactions between the atoms \cite{Staniscia,Schuetz:2014,Tesio:2014}, similarly to gravitational and Coulomb potential in two or more dimensions \cite{Campa:2009}. In view of this analogy, it is relevant to study the dynamics of these systems at and out-of-equilibrium so to test in a laboratory conjectures and predictions, such as ensemble inequivalence and the existence of quasi-stationary states \cite{Staniscia}. %In the study of these systems, in particular, equilibrium properties are well described by mean-field models \cite{Campa:2009}. This urges one to systematically investigate in how far this is also applicable to the collective dynamics of cold atoms in standing-wave resonators.  
The realization in quantum optical setups, like the one sketched in Fig. \ref{fig:1}(a), brings additional peculiar features. In fact, these systems are intrinsically lossy, so that non-trivial dynamics can be observed only in presence of a pump. On the one hand, the conservative potential mediated by the cavity photons shares several analogies with the one of the Hamiltonian-Mean-Field model \cite{HMF,Schuetz:2014,Campa:2009, Schuetz:2015}, of which several features are well reproduced by a mean-field description  \cite{HMF,Campa:2009}.   On the other hand, cavity losses give rise to damping and diffusion, which are characterized by a spatial structure, thus establishing long-range correlations between the atoms  \cite{Schuetz:2013,Schuetz:2014}. These correlations, in turn, cannot be captured by a mean-field description. \\
\indent In this work we systematically develop a mean-field model for cold atoms in a standing-wave resonator, in the setup illustrated in Fig. \ref{fig:1}(a), and test its validity by comparing its predictions with the ones of the Fokker-Planck equation for the full atoms distribution \cite{Schuetz:2013}. %This comparison also allows us to analyse the effects of the dissipative long-range forces on the atoms dynamics.
This work completes a series of papers, which analyse the equilibrium and out-of-equilibrium dynamics of spatial self-organization of atomic ensembles in a single-mode resonator.  Our analysis is based on a semiclassical treatment, and specifically on a Fokker-Planck equation (FPE) for the $N$-atom distribution, derived when the atoms are classically polarizable particles and  their center-of-mass motion is confined to one dimension \cite{Schuetz:2013}. The cavity field, instead, is a full quantum variable. This makes our treatment applicable also in the shot-noise limit \cite{Schuetz:2013} and gives access to regimes that are complementary to those based on the model in Ref. \cite{Horak:2001}, where the field is a semiclassical variable. \\
Our formalism permits us to consistently eliminate the cavity variables from the equations of motion of the atoms and to investigate the properties of the cavity field across the self-organization threshold, where the intracavity field is characterized by large fluctuations. Starting from this model in Ref. \cite{Schuetz:2015} we analysed the stationary state of the $N$-body FPE, and showed that (i) this is a thermal state whose temperature is determined by the linewidth of the resonator, and (ii) that the transition to self-organization is a Landau-type second-order phase transition, as illustrated in Fig. \ref{fig:1}(b)-(c). In Ref. \cite{Schuetz:2015} we also determined the corresponding phase diagram as a function of the physical parameters and predicted the corresponding features in the light emitted by the resonator. In Ref. \cite{Schuetz:2015b} we investigated the dynamics following sudden quenches across the phase transition, and found that the interplay between long-range conservative and dissipative forces gives rise to prethermalization dynamics, where the long-range nature of dissipation plays an essential role. \\
In this work we derive a mean-field treatment from our $N$-atom FPE. We then benchmark the limits of validity of the mean-field ansatz by means of numerical simulations using the full $N$-body FPE and by means of a BBGKY hierarchy. The results we obtain are compared with existing literature on spatial self-organization in single-mode cavities, both for the semiclassical treatment  \cite{Asboth:2005,Niedenzu:2011,Griesser:2012,DallaTorre:2013}, as well as for the case in which the atomic quantum statistics is assumed to be relevant \cite{Nagy:2010,Nagy:2011,Piazza:2013,Oeztop:2012,Kulkarni:2013,DallaTorre:2013,Piazza:2014}.

This work is organized as follows. In Sec. \ref{Sec:Model} the Fokker-Planck Equation at the basis of our analysis is reported and the corresponding mean-field equation is derived. In Sec. \ref{Sec:Steadystate} the stationary properties of the mean-field FPE distribution function are analytically determined. The mean-field predictions are compared with the ones of the $N$-body FPE and with further existing theoretical works.  In Sec. \ref{Sec:Vlasov} the Vlasov equation, which describes the short time dynamics of the mean-field FPE, is derived. Its predictions are then determined by means of a stability analysis and the analytical results are compared with the numerical simulations of the mean-field FPE. Section \ref{Sec:Limits} reports a critical analysis of the limits of validity of the mean-field treatment. In Sec. \ref{Sec:Conclusions} the conclusions are drawn, while in the Appendix \ref{AppendixA} calculations are reported that complement the material presented in Sec. \ref{Sec:Steadystate}. 

\section{Derivation of the mean-field model}
\label{Sec:Model}

In this section we derive the mean-field model starting from the Fokker-Planck equation (FPE) describing the dynamics of an atomic ensemble in the optical potential of a high-finesse resonator of Ref. \cite{Schuetz:2013}. The atoms are $N$, have mass $m$, their motion is assumed to be confined along the $x$-axis, which also coincides with the axis of a high-finesse cavity and within whose mirrors the atoms are spatially trapped. In the following we denote their canonically-conjugated positions and momenta by $x_j$ and $p_j$ ($j=1,\ldots,N$). The atomic dipole strongly couples to one cavity mode and is transversally driven by a laser, as sketched in Fig. \ref{fig:1}(a). The parameter regime is such that the atoms coherently scatter photons into the cavity mode  and their external motion is determined by the light forces associated with these processes. The light forces are periodic, and their period is determined by the cavity mode standing wave, whose spatial mode function is $\cos (kx)$, with $k$ the cavity-mode wave number.

 \subsection{Basic assumptions}
 
Before reporting the FPE which governs the dynamics of the $N$-body distribution function, we summarize the main approximations behind its derivation and the corresponding physical parameters.

One basic assumption of our model is that the only relevant scattering processes are coherent. This regime can be reached when the cavity mode and laser frequencies are tuned far off resonance from the atomic transition \cite{Horak:1997,Vuletic:2000}. We denote by  $\Delta_a=\omega_L-\omega_0$ the detuning between laser ($\omega_L$) and atomic frequency ($\omega_0$), and assume that this is the largest parameter of the problem. It is thus larger than the coupling strengths for the interaction between dipole and fields. It is also larger than the detuning  $\Delta_c=\omega_L-\omega_c$ between laser and cavity mode frequency, whose wave numbers are to good approximation denoted by the same parameter $k$. This allows us to eliminate the internal degrees of freedom of the atoms by a perturbative expansion in the lowest order of the small parameter $1/|\Delta_a|$.  

The cavity field is treated as a quantum mechanical variable and the dynamics can be cast as an opto-mechanical coupling between atomic motion and cavity field \cite{Horak:2001,Domokos:2002}.  The parameter regime we assume gives rise to a time-scale separation, such that the cavity degrees of freedom evolve on a faster time-scale than the motion. This is warranted when the cavity line width $\kappa$, which determines the relaxation rate of the resonator state, is much larger than the recoil frequency $\omega_r=\hbar  k^2/(2m)$, which scales the exchange of mechanical energy between light and atoms. In this limit the cavity field is eliminated from the equations of motion of the atomic external degrees of freedom in a perturbative expansion to first order in the small parameter $1/\kappa$, implementing a procedure first applied in Ref. \cite{Dalibard:1985}. The hierarchy of time scales is set by the inequalities $|\Delta_a|\gg\kappa\gg\omega_r$. This is also consistent with a semiclassical treatment, since the kinetic energy of the atoms at steady state scales with $\hbar\kappa$ thus warranting that the width  $\Delta p$ of the single-atom momentum distribution is large in comparison to the linear momentum $\hbar k$ carried by each photon \cite{Schuetz:2014,Schuetz:2015,Domokos:2002}. 

 \subsection{Collective motion of $N$ atoms in a cavity field}

The approximations above discussed are at the basis of the theoretical procedure which connects the master equation of atoms in a quantized cavity field with the FPE for the Wigner function $f_N=f_N(x_1,\ldots, x_N;p_1,\ldots, p_N ;t)$, describing the positions and momenta of the $N$ atoms at time $t$. The derivation is detailed in Ref. \cite{Schuetz:2013} and the resulting FPE reads
\begin{align}
\frac{\partial f_N}{\partial t}=&-\sum_{i=1}^{N}\frac{\partial}{\partial x_i}\frac{p_i}{m}f_N+S^2L[f_N]\label{FPE}\,,
\end{align}
where the second summand on the right-hand side (RHS) is due to mechanical effects of the cavity field on the atoms and scales like $S^2$. 
Here $S=\Omega g/\Delta_a$ is the scattering amplitude between laser and cavity mode, it is proportional to the laser strength $\Omega $ and to the cavity vacuum Rabi frequency $g$, which scale the interaction between dipole and laser and between dipole and cavity, respectively. Operator $L[f_N]$ takes the form
\begin{subequations}
\begin{align}
S^2L[f_N]=
&%- \sum_i \frac{\partial}{\partial p_i} F_0 N\Theta(x_1,\ldots,x_N)\sin(kx_i)f_N \label{force}\\  
\frac{\partial f_N}{\partial p_i}\frac{\partial V(x_1,\ldots,x_N)}{\partial x_i}\\
&- S^2\sum_{i,j}  \frac{\partial}{\partial p_i} \Gamma_0 \sin(kx_i) \sin(kx_j) p_jf_N 
\label{friction}\\
&+S^2\sum_{i,j}  \frac{\partial^2}{\partial p_i \partial p_j}D_0 \sin(kx_i) \sin(kx_j)f_N
\label{diffusion}\\
&+ S^2 \sum_{i,j} \frac{\partial^2}{ \partial p_j\partial x_i}\eta_0\sin(kx_i) \sin(kx_j)f_N\,.
\label{etaterm}
\end{align}
\label{L:f}
\end{subequations}
Each line on the RHS of Eq. \eqref{L:f} has a physical meaning. The first term describes the dynamics due to the conservative potential
\begin{align}
V(x_1,\ldots,x_N)=\frac{\hbar\Delta_c}{\kappa^2+\Delta_c^2}S^2N^2\Theta(x_1,\ldots,x_N)^2%=\frac{F_0N^2S^2}{2k}\Theta^2
\,,
\label{potential}
\end{align}
where
\beq
\label{Theta:N}
\Theta(x_1,\ldots,x_N)=\frac{1}{N}\sum_{j=1}^N\cos(kx_j)\,,
\eeq
so that the potential mediates long-range interactions between the atoms. Parameter $\langle|\Theta|\rangle_N$ is the order parameter of self-organization, where $\langle \cdot \rangle_N$ denotes the expectation value taken over the normalized distribution $f_N$. Specifically, when the atoms form Bragg grating, then $\langle|\Theta|\rangle_N\to 1$ and the potential depth is maximal. When the atoms are instead uniformly distributed in space, then $\langle|\Theta|\rangle_N\simeq 0$ and the potential vanishes. We note that the Bragg gratings minimize the potential when $\Delta_c<0$, otherwise the uniform distribution is energetically favoured. We will here denote $\langle|\Theta|\rangle_N$ by magnetization, due to the mapping of the self-organization transition to a ferromagnetic model \cite{Schuetz:2015}.\\
For later convenience, we define the parameter 
\begin{align}
F_0 = (\hbar k)  \frac{2 \Delta_c}{\kappa^2+\Delta_c^2}\,,
\label{F}
\end{align}
such that $V=F_0(NS\Theta)^2/(2k)$.

The second term on the RHS, Eq. \eqref{friction}, describes a dissipative force and is scaled by the coefficient $\Gamma_0$:
\begin{align}
\Gamma_0=\omega_r  \frac{8 \Delta_c \kappa }{(\kappa^2+\Delta_c^2 )^2}\,.
\label{Gamma}
\end{align}
This term is due to non-adiabatic corrections in the dynamics of the cavity field. 

Term in the line \eqref{diffusion} corresponds to diffusion due to fluctuations of the cavity field associated with losses.  
 The diffusion matrix is the dyadic product of the vector $(\sin(kx_1),...,\sin(kx_N))$ with itself and scales with the coefficient 
\begin{align}
D_0=(\hbar k)^2 \frac{ \kappa }{\kappa^2 + \Delta_c^2}\,.
\label{D}
\end{align} Therefore, beside the diffusion due to the diagonal elements, which is a single-particle effect, we also expect that term \eqref{diffusion} establishes long-range correlations. 

The last line \eqref{etaterm} contains cross-derivatives and scales with the coefficient
\begin{align}
 \eta_0=2\hbar \omega_r \frac{\kappa^2-\Delta_c^2}{(\kappa^2 + \Delta_c^2)^2}\,,
 \label{eta}
\end{align}
whose sign depends on whether the ratio $|\Delta_c/\kappa|$ is smaller or larger than unity, while it vanishes for $|\Delta_c/\kappa|=1$. An analogous term has also been reported in the semiclassical description of cold atoms in optical lattices \cite{Dalibard:1985}, where it has been then neglected under the assumption of uniform spatial densities. Such assumption cannot be applied in the self-organized regime, nevertheless we will show that this term can be consistently discarded in the thermodynamic limit we apply, which warrants Kac's scaling \cite{Campa:2009}. 

\subsection{Mean-field ansatz}
\label{Sec:meanFPE}

To derive a mean-field FPE we assume that the Wigner function is factorized into single-particle distribution functions according to the prescription
\begin{align}
f_N(x_1,\ldots,x_N;p_1,\ldots,p_N;t)=\prod_{i=1}^{N}f_1(x_i,p_i;t)\,,
\label{factorization}
\end{align}
where $f_1(x_i,p_i;t)$ denotes the distribution for the particle $i$ at time $t$ and is thus defined on the phase space of this particle. We use then Eq. \eqref{factorization} in the FPE \eqref{FPE} and integrate out all particles' variables but one. In this way we derive the mean-field FPE, which reads
\begin{align}
\frac{\partial f_1}{\partial t}&=-\frac{\partial }{\partial x}\frac{p}{m}f_1+S^2\mathfrak{L}[f_1]\,,
\label{meanFPE}
\end{align}
and has same structure as the FPE in Eq. \eqref{FPE}. Operator $\mathfrak{L}$ describes, as  $L$, the mechanical effects of light. However, it is now a nonlinear operator of $f_1$ and takes the form
\begin{subequations}
\begin{align}
\mathfrak{L}[f_1]=&-\frac{\partial}{\partial p} F_0\left(\cos(kx)+(N-1)\Theta_{\mathrm{MF}}[f_1]\right)\sin(kx)f_1 \label{meanforce}\\  
&- \frac{\partial}{\partial p} \Gamma_0\left(\sin(kx)p+(N-1)\Xi_{\mathrm{MF}}[f_1]\right)\sin(kx)f_1 
\label{meanfriction}\\
&+\frac{\partial^2}{\partial^2 p }D_0 \sin^2(kx)f_1\label{meandiffusion}\\
&+  \frac{\partial^2}{ \partial p \partial x}\eta_0\sin^2(kx)f_1\,,\label{meaneta}
\end{align}
\label{meanL:f}
\end{subequations}
where we have introduced the functionals
\begin{align}
\Theta_{\mathrm{MF}}[f_1]&=\frac{1}{\lambda}\int_{0}^{\lambda}dx\int_{-\infty}^{\infty}dp\cos(kx)f_1\,,\label{meanTheta} \\ 
\Xi_{\mathrm{MF}}[f_1]&=\frac{1}{\lambda}\int_{0}^{\lambda}dx\int_{-\infty}^{\infty}dp\sin(kx)pf_1\,. \label{meanXi}
\end{align}
The mean-field order parameter $\Theta_{\rm MF}$ is the expectation value $\langle \cos(kx)\rangle$, where $\langle . \rangle$ indicates the average taken over the single-particle distribution function $f_1(x,p)$. 
%The functi?onal $\Theta[f_1]$ is the mean-field value of the $N$-atom function $\Theta(x)$ and this is the reason why it gets the same name here. The second functional $\Xi[f_1]$ is the part of \eqref{friction} where the force is also depending on the velocities of the other particles. In both cases the functionals give the interaction of the atoms. 
The terms on the RHS contained in lines \eqref{meanforce} and \eqref{meanfriction} have a different origin but a similar structure, which can be recognized by analysing the form of the two summands within the respective inner brackets. The first summand in each line describes the interaction of the atom with itself, mediated by the cavity field. The second summand in each line emerges from the interaction between the atom and all other $N-1$ atoms. 

We further notice that the term in line \eqref{meanforce} can be cast in terms of a conservative force originated from the potential
\begin{align}
V_1[f_1](x)=&\frac{F_0}{2k}S^2\left(\cos^2(kx)+2(N-1)\Theta_{\mathrm{MF}}[f_1]\cos(kx)\right)\nonumber\\
&+\frac{\Gamma_0}{k}(N-1)S^2\Xi_{\mathrm{MF}}[f_1]\cos(kx)\,,
\label{meanpotential}
\end{align}
and contains a term, whose corresponding term in Eq. \eqref{FPE} has dissipative nature (see line \eqref{friction}).
%whereby the first line in \eqref{meanpotential} is just the analogue to \eqref{potential} and the second line is due to the off-diagonal friction. 
Using this result, we can rewrite Eq. \eqref{meanL:f} in the compact form
\begin{align*}
\mathfrak{L}[f_1]=\frac{\partial V_1}{\partial x}\frac{\partial f_1}{\partial p}- \frac{\partial}{\partial p} \left(\Gamma_0p -\frac{\partial}{\partial p }D_0-\frac{\partial}{\partial x}\eta_0\right)\sin^2(kx)f_1 \,,
\end{align*}
which allows us to simply read out the physical meaning of the other terms, they are in fact the diagonal component of friction, diffusion, and cross-derivative term in Eq. \eqref{FPE}.
%In the following sections we discuss the predictions of the mean-field FPE. %Being this, however, an equation based on ansatz which discards particle-particle correlations, it has a limited validity, which we will discuss in Sec. \ref{Sec:Validity}.

\section{Stationary state of the mean-field equation}
\label{Sec:Steadystate}

The stationary properties of the mean-field distribution are analysed by means of the single-particle distribution $f_{\text{st}}(x,p)$ that solves Eq. \eqref{meanFPE} with
\begin{equation}
\label{eq:steady}
\partial_tf_{\text{st}}(x,p)=0\,.
\end{equation} 
In the following we determine $f_{\text{st}}(x,p)$ and then analyse its predictions for relevant physical quantities.

\subsection{Derivation of the steady state solution}

In order to solve Eq. \eqref{eq:steady} we consider the ansatz
\begin{align*}
f_{\text{st}}(x,p)=f_0\exp(a(x)+b(p))\,,
\end{align*}
where $a(x)$ and $b(p)$ are functions which only depend on position and momentum, respectively, and $f_0$ is the normalization constant. Using this ansatz in Eq. \eqref{meanFPE} we obtain differential equations for $a(x)$ and $b(p)$, whose solutions read $b(p)=-\beta p^2/(2m)$ and 
\begin{align}
\label{a}
a(x)&=(Y/2-1)\ln(1+Z\sin^2(kx))\\
 &-(N-1)Y\Theta_{\mathrm{MF}}[f_{\text{st}}]\sqrt{\frac{Z}{1+Z}}\mathrm{arctanh}\left(\sqrt{\frac{Z}{1+Z}}\cos(kx)\right)\,,\nonumber
\end{align}
with $Y=F_0/(k\eta_0)$, $Z=\beta\eta_0 S^2$, and
%\begin{align}
%a=&\left(\frac{F_0}{2k\eta_0}-1\right)\ln\left(1+\beta\eta_0S^2\sin^2(kx)\right)\nonumber\\
%&-\frac{F_0(N-1)S^2\beta}{k}\Theta[f_{\text{st}}]\frac{\tanh^{-1}\left(\frac{\sqrt{\beta\eta_0S^2}}{\sqrt{\beta\eta_0S^2+1}}\cos(kx)\right)}{\sqrt{\beta^2\eta_0^2S^4+\beta \eta_0S^2}},\label{a}\\
%\end{align}
%The constant $\beta$ is given by
\begin{align}
\beta=-\frac{\Gamma_0 m}{D_0}=\frac{-4\Delta_c}{\hbar\left(\kappa^2+\Delta_c^2\right)}\,.
\label{finaltemperature}
\end{align}
Therefore, 
\begin{align}
\label{steady:1}
f_{\text{st}}(x,p)=\mathcal F(\cos kx)\exp\left(-\beta\frac{p^2}{2m}\right)\,,
\end{align}
with $\mathcal F(\cos kx)=f_0\exp(a(x))$. Equation \eqref{steady:1} describes a thermal distribution provided that $\Delta_c<0$: In this limit parameter $\beta$, Eq. \eqref{finaltemperature}, plays the role of an inverse temperature at steady state. This temperature coincides with the value found by solving the steady state of the $N$-body FPE, Eq. \eqref{FPE}, as shown in Refs. \cite{Schuetz:2014,Schuetz:2015}. 

We note that the function $\mathcal F(\cos k x)$ depends on $\Theta_{\mathrm{MF}}[f_{\text{st}}]$, which leads to the fixed-point equation
\begin{align}
\Theta_{\mathrm{MF}}\equiv\left\langle\cos(kx) \right \rangle=\sqrt{\frac{2\pi m}{\beta}}\frac{1}{\lambda}\int_{0}^{\lambda}{\rm d}x\cos(kx)\mathcal F(\cos kx)\,.
\label{fixpoint}
\end{align} 
Its solution is in general not transparent, but it gets simpler in an appropriately defined thermodynamic limit. This consists in scaling the coupling strength $g\sim 1/\sqrt{N}$ as the number of atoms is increased, leading to the scaling relation $S\propto 1/\sqrt{N}$ \cite{Larson:2008,Fernandez-Vidal:2010}. In this limit function $a(x)$, Eq. \eqref{a}, can be cast into the form 
\begin{align}
a(x)=2\frac{\bar{n}}{\bar{n}_c}\Theta_{\mathrm{MF}} \cos(kx)
\label{anew}
\end{align}
with 
\begin{align}
\bar{n}=\frac{NS^2}{\kappa^2+\Delta_c^2} \label{bar:n}\,,
\end{align}
and
\begin{align}
\bar{n}_c=\frac{\kappa^2+\Delta_c^2}{4\Delta_c^2} \label{threshold}\,.
\end{align}
This leads to a compact form of the stationary distribution in the mean-field limit:
\begin{align}
f_{\text{st}}(x,p)&=f_0\exp\left(-\beta\left(\frac{p^2}{2m}+\hbar\Delta_c\bar{n}\Theta_{\mathrm{MF}}\cos(kx)\right)\right)\,,
\label{f:st}
\end{align}
with 
\begin{align*}
f_0^{-1}=\sqrt{\frac{2m\pi}{\beta}}I_0\left(2\frac{\bar{n}}{\bar{n}_c}\Theta_{\mathrm{MF}}\right)\,,
\end{align*}
and $I_j$ is the modified Bessel function of $j$-th order \cite{Abramowitz}. 

We thus see that in the thermodynamic limit the effect of the cross derivatives vanishes. For finite $N$, parameter $\eta_0$ is small but finite and in the stationary state it gives rise to a correction to the effective potential term, as visible in Eq. \eqref{a}.

\begin{figure}
	\includegraphics[width=1\linewidth]{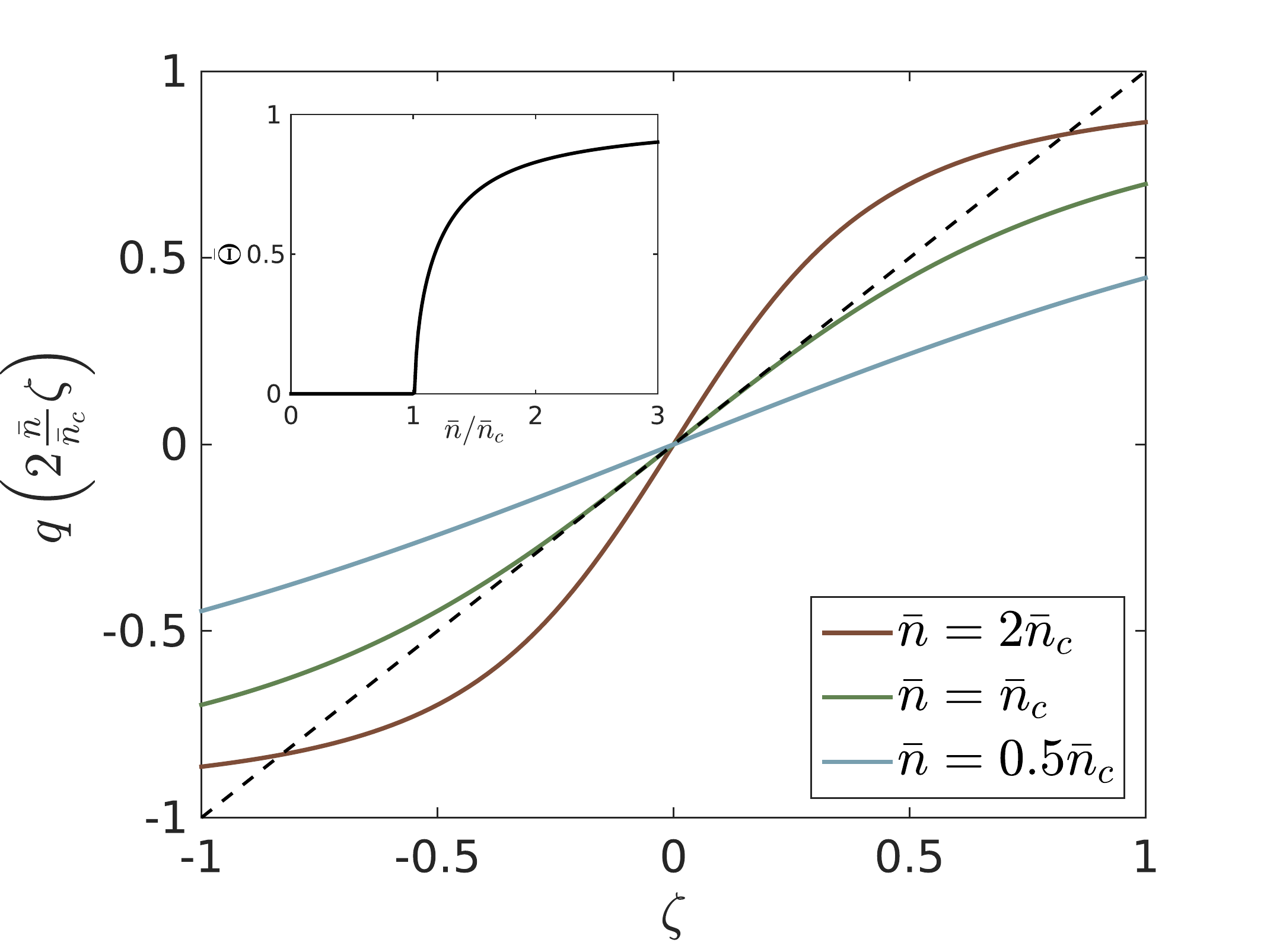}
	\caption{(Color online) Onset: Plot of  $q(2\bar n \zeta /\bar n_c)$, Eq. \eqref{f:q}, as a function of $\zeta$ and for different values of $\bar n$. The intersection points with the curve $y=\zeta$ (dashed line) give the solutions of Eq.  \eqref{fixpointnew}. Stable points are at the crossing where $q'<\bar{n}_c/(2\bar{n})$ and are the equilibrium values of the order parameter $\Theta_{\mathrm{MF}}$. Inset: The resulting stable solution $\bar{\Theta}\geq0$ as a function of $\bar n$ (in units of $\bar n_c$).\label{fig:2}}
\end{figure}

\subsection{Stationary properties in the thermodynamic limit} 

The mean-field distribution, Eq. \eqref{f:st}, allows one to analytically determine several properties of the steady state. First, functional $\Theta_{\mathrm{MF}}$ in the exponent has to be determined self-consistently. Using Eq. \eqref{anew} in Eq. \eqref{fixpoint} gives the relation
\begin{align}
\label{fixpointnew}
\Theta_{\mathrm{MF}}=q\left(2\frac{\bar{n}}{\bar{n}_c}\Theta_{\mathrm{MF}}\right)\,,
\end{align}
where $q$ is the function of the form 
\begin{align}
q\left(2\frac{\bar{n}}{\bar{n}_c}\zeta\right)=\frac{I_{1}\left(2\frac{\bar{n}}{\bar{n}_c}\zeta\right)}{I_{0}\left(2\frac{\bar{n}}{\bar{n}_c}\zeta\right)}\,,
\label{f:q}
\end{align}
and is plotted in Fig. \ref{fig:2} for values of $\bar n$ below, at, and above $\bar n_c$. The solutions of Eq. \eqref{fixpointnew} are the crossing between the curve $y=\zeta$ and $y=q\left(2\bar{n}\zeta/\bar{n}_c\right)$, see Eq. \eqref{f:q}.  For $\bar n<\bar n_c$ 
this equation allows for one solution, corresponding to $\Theta_{\mathrm{MF}}=0$. For $\bar n>\bar n_c$, the solutions are three, of which two are stable and one is unstable. The stable solutions give $\Theta_{\mathrm{MF}}=\pm \bar{\Theta}$, with $0\leq\bar{\Theta}< 1$, and correspond to the self-organized state. Close, but above, the critical point the value $\bar{\Theta}$ can be analytically determined and reads
\begin{equation}
\label{Theta:0}
\bar{\Theta}=\sqrt{2(\bar n/\bar n_c-1)}\,.
\end{equation}
The value $\bar{n}=\bar{n}_c$, with $\bar{n}_c$ defined in Eq. \eqref{threshold}, determines hence a critical point, at which the transition to self-organization occurs, and is controlled by the detuning from the cavity field and the cavity loss rate, for the thermodynamic limit we chose. The results we obtained so far for the stationary mean-field distribution are in full agreement with the ones found for the stationary distribution of Eq. \eqref{FPE}, see Ref.  \cite{Schuetz:2015}. The stationary mean-field distribution in Eq. \eqref{f:st} corresponds to the one that is found from the stationary $N$-particle distribution after integrating out the other $N-1$ position and momentum variables, and then taking the thermodynamic limit. The equation for the order parameter, Eq. \eqref{fixpointnew}, agrees with the one obtained for the $N$-particle case and obtained by means of a saddle-point approximation. This agreement is found also for the critical value of Eq. \eqref{threshold} and for the temperature of Eq. \eqref{finaltemperature}. Hence, the mean-field model predicts the same phase diagram as the $N$-body FPE.

It is also instructive to consider the value of the bunching parameter $\mathcal{B}$ as a function of $\bar n$. This is defined as
\begin{align}
\mathcal{B}=\left\langle\cos^2(kx) \right \rangle\,,
\label{bunching}
\end{align}
and gives a measure of localization of the particles at the minima of the mechanical potential \cite{Schuetz:2015, Asboth:2005}. Using Eq. \eqref{anew} we obtain 
\begin{align}
\mathcal{B}=
\begin{cases}
1/2\,,&\bar{n}\leq \bar{n}_c,\\
1-\bar{n}_c/(2\bar{n})\,,& \bar{n}>\bar{n}_c\,,
\end{cases}
\label{bunchingcases}
\end{align}
in the stationary state.
Therefore, below threshold the atoms are uniformly distributed, while above threshold they increasingly localize at the minima of the Bragg potential. In particular, when the atoms are tightly-bound at the minima, the above-threshold expression in Eq. \eqref{bunchingcases} delivers the amplitude of the fluctuations, namely,
\begin{equation}
k^2\langle x^2\rangle\approx \frac{\bar n_c}{2\bar n}\,,
\end{equation}
showing that these are inversely proportional to the laser intensity. 

\subsection{Comparison with existing literature}

The results obtained so far by means of the mean-field model show a remarkable agreement with the predictions of the stationary solution of the $N$-particle FPE, Eq. \eqref{FPE}. It is further worthwhile to compare the results here derived with the results obtained in the literature by means of different approaches. %This sheds some light into the links between the various models and the explicit and hidden assumptions. 

We first discuss Ref. \cite{Asboth:2005}, where, amongst other studies, a mean-field approach is developed based on plausible conjectures. Here, the mean-field potential is calculated and the threshold of self-organization is determined by (i) assuming that the stationary state is thermal, with temperature given by the linewidth of the cavity, and (ii)  performing a stability analysis of the uniform density distribution. By means of this study a threshold value for self-organization is identified, which agrees with the prediction in Eq. \eqref{threshold}, as it becomes evident after defining the threshold amplitude $S_c$ such that 
\begin{align*}
\frac{NS_c^2}{\Delta_c^2+\kappa^2}\equiv\bar{n}_c\,.
\end{align*}
In particular, the quantity $\eta^*$ in \cite{Asboth:2005} is in our notations $S_c\Delta_a/g$ calculated for the case $\Delta_c=-\kappa$.

The stationary state of self-organization has been first derived in the following works \cite{Niedenzu:2011,Griesser:2012} by means of a FPE as a function of the atomic and field variables. This description assumes that the field fluctuations are small, and thus cannot reliably reproduce the field correlation functions below and at threshold. It predicts, nevertheless, that the atoms steady state is thermal and its temperature coincides with the inverse of Eq. \eqref{finaltemperature}, apart for corrections of the order $\omega_r/\kappa$, that are systematically neglected in our approach because they are of higher order. It further predicts the same behaviour of the order parameter as in Eq. \eqref{Theta:0} above, but close, to threshold.

It is also interesting to compare our results with a series of other theoretical studies, which focus on self-organization of ultracold atomic ensembles in cavities but discard retardation effects: In these works only the conservative part of the cavity potential is considered,  while the temperature at steady state is due to the coupling to an external heat bath \cite{Nagy:2010,Nagy:2011,Piazza:2013,Oeztop:2012,Kulkarni:2013,DallaTorre:2013,Piazza:2014}. Even though the conditions seem quite different from our case, remarkable agreement is found in the appropriate limits. References  \cite{Nagy:2010,Nagy:2011} analyse the self-organization transition of an ultracold gas of bosonic atoms and derive the mapping to the Dicke model. Here, the recoil energy plays an analogous role as the temperature, and the threshold which is derived agrees with the threshold in Eq. \eqref{threshold} after setting 
\begin{align}
NS_c^2=\frac{1}{\beta}\frac{\kappa^2+\Delta_c^2}{-\Delta_c}\label{Sc}\,,
\end{align}
with $\beta=4/\hbar\omega_r$. By means of this prescription, the threshold also agrees with the one calculated in Ref. \cite{DallaTorre:2013}. Furthermore, it also coincides with the one evaluated in Ref. \cite{Piazza:2014} when using the Boltzmann distribution for the atoms statistics. \\
Another quantity which has been determined in these works is the photon flux, which corresponds to the intracavity photon number in our treatment. In Refs. \cite{Nagy:2011,Oeztop:2012,Kulkarni:2013,DallaTorre:2013} the photon flux scales as $1/|\bar n-\bar n_c|$ below threshold, while at threshold it diverges as $\sqrt{N}$. These predictions are in perfect agreement with the results we find taking the stationary distribution of Eq. \eqref{FPE}, see Appendix \ref{AppendixA}, Eqs. \eqref{n:cav:below} and \eqref{n:cav:at}. In particular, the intracavity photon number at threshold, Eq. \eqref{n:cav:at}, coincides with the one calculated in Ref.  \cite{DallaTorre:2013} after substituting in their equation $\omega_z=(\omega_0^2+\kappa^2)/\omega_0$ for the temperature, with $\omega_0=-\Delta_c$. The result for the intensity-intensity correlations at zero-time delay and below threshold, Eq. \eqref{g:2:below}, further agrees with the result derived in Ref. \cite{Oeztop:2012,Kulkarni:2013}. 

%The general agreement between the results of the semiclassical mean-field theory and various treatments, based on different theoretical approaches, seem to hint that such approaches also discard effects beyond mean-field. It also suggests that the mean-field treatment is a valid tool, which is moreover relatively simpler, for extracting estimates about the stationary properties of the selforganized state in the quantum regime. 

\section{Mean-field Dynamics}
\label{Sec:Vlasov}

We now study the dynamics predicted by the mean-field FPE. We focus on the Vlasov equation, which we derive from Eq. \eqref{meanFPE} by taking the thermodynamic limit, according to our prescription. The Vlasov equation for our problem reads
\begin{align}
\frac{\partial f_1}{\partial t}+\frac{p}{m}\frac{\partial f_1}{\partial x_1}-\frac{\partial V_0[f_1](x)}{\partial x}\frac{\partial f_1}{\partial p}=0
\label{vlasov}
\end{align}
with
\begin{align}
V_0[f_1](x)=&2\hbar\Delta_c\bar{n}\cos(kx)\Theta_{\mathrm{MF}}[f_1]\nonumber\\
&-\frac{\hbar^2 k}{m}\bar{n}\beta\kappa\cos(kx)\Xi_{\mathrm{MF}}[f_1]\label{meanpotentialnew}\,,
\end{align}
and it corresponds to the potential in Eq. \eqref{meanpotential} after neglecting the self-reaction term, which is of order $1/N$. Therefore, the validity of the predictions we will extract are limited to sufficiently short time scales for which the corrections can be discarded. We will quantify this statement in the next section. 

%In this section we describe the dynamics in the mean-field. We start with the Vlasov equation that is the rigorous mean-field description of the whole FPE. Rigorous means here that it describes the FPE on timescales of the order $N$ neglecting correction with the order of magnitude $1/N$. This Vlasov equation does not describe the conservation of the energy of the system because of mean-field friction effects. The second part would be to describe a cooling rate in the mean-field. We see that there is some fixed cooling rate when the system is organized whereby the cooling rate increases with $S$ when the system is homogeneous.

\subsection{Preliminary considerations: energy conservation}

We first analyse whether  Eq. \eqref{vlasov} warrants energy conservation. We consider a class of functions for which $\Xi_{\mathrm{MF}}[f_1]=0$. This includes the stationary solution of Eq. \eqref{f:st}. For these solutions, the energy of one particle takes the form
\begin{align}
\label{energy:vlasov}
\epsilon(t)=\frac{\left\langle p^2\right \rangle }{2m}+\hbar\Delta_c\bar{n}\Theta_{\mathrm{MF}}^2\,.
\end{align}
In order to determine $\dot\epsilon(t)$ we thus calculate $\dot{\Theta}_{\mathrm{MF}}$ and $\dot{\langle p^2 \rangle}$. This gives
\begin{align*}
\dot{\Theta}_{\mathrm{MF}}&=-\frac{k}{m}\Xi_{\mathrm{MF}}\,,\\
\frac{\dot{\langle p^2 \rangle}}{2m}&=2\frac{\hbar}{m} \bar{n}\left(k\Delta_c\Theta_{\mathrm{MF}}-\omega_r\kappa\beta\Xi_{\mathrm{MF}}\right)\Xi_{\mathrm{MF}}\,,
\end{align*}
and  therefore we get for the derivative of the energy
\begin{align*}
\dot{\epsilon}=-2\frac{\hbar}{m} \bar{n}\omega_r\kappa\beta\Xi_{\mathrm{MF}}^2\,.
\end{align*}
These derivatives hence vanish when $\Xi_{\mathrm{MF}}=0$, and thus for the class of distribution fulfilling this condition, energy, with the potential term given in Eq. \eqref{energy:vlasov}, is conserved. Fluctuations, on the other hand, can give rise to finite values of $\Xi_{\mathrm{MF}}$. The purpose of the next section is to analyse the stability and short-time dynamics of solutions of the Vlasov equation, Eq. \eqref{vlasov}, after quenches of the laser parameters.

\subsection{Stability analysis of spatially homogeneous distributions}
\label{Sec:Stability}

We now analyse the short-time dynamics described by Eq. \eqref{vlasov}, assuming that at $t=0$  the distribution is thermal and with uniform spatial density, thus $f_1(x,p,0)=f_1(p,0)$ and $\left. \Theta_{\mathrm{MF}}\right|_{t=0^-}=0$, with 
\begin{align}
f_1(p,0)=\left(\frac{2m\pi}{\beta_0}\right)^{-\frac{1}{2}}\exp\left(-\beta_0\frac{p^2}{2m}\right)\,,\label{meanf0}
\end{align}
where $\beta_0$ is the inverse temperature. This distribution is a stable solution of the Vlasov equation after setting $\bar n=0$. At $t=0$ the laser strength is quenched above threshold so that parameter $\bar n$ takes a finite value larger than $\bar n_c$. We then let evolve the distribution of Eq. \eqref{meanf0}  by taking this value $\bar n$ in Eq. \eqref{vlasov}. 
Figure \ref{vlasovplot} shows the results of the numerical integration of Eq. \eqref{vlasov}  for different value of $\bar n$. We analyse these results, keeping in mind that they are strictly valid for short times since the Vlasov equation discards effects, such as diffusion, which are crucial in determining the stationary state. In (a)  the order parameter evolves from 0 to a finite value, about which it oscillates. This value is smaller than the one predicted by the stationary solution of the mean-field FPE. It is reached after an initial dynamics characterized by an exponential increase, whose slope is steeper the larger is $\bar n$. Subplots (b) and (c) display the corresponding evolution of the quantities $\Xi_{\mathrm{MF}}^2$, see Eq. \eqref{meanXi}. This quantity emerges from the retardation effects of the dynamics, it is thus a signature of memory effects, and mathematically corresponds to the build up correlations between momentum and position that cannot be factorized. The initial distribution, Eq. \eqref{meanf0}, is chosen so that $\Xi_{\mathrm{MF}}=0$, and we observe that the dynamics give rise to a build up of a finite value of $\Xi_{\mathrm{MF}}^2$, with an exponential increase that leads to a maximum where the curve for $\Theta_{\mathrm{MF}}$ reaches the plateau. Then, it oscillates like $\Theta_{\mathrm{MF}}$ (one can well understand the behaviour of these oscillations observing that $\Xi_{\mathrm{MF}}$ is proportional to the time derivative of $\Theta_{\mathrm{MF}}$) and is exponentially damped to zero. In the initial phase, the exponential growth of $\Xi_{\mathrm{MF}}^2$ increases with $\bar n$, similarly in the second phase of the dynamics, where $\Theta_{\mathrm{MF}}$ oscillates about a finite mean value, the amplitude of the oscillations of  $\Xi_{\mathrm{MF}}^2$ are also larger the larger is $\bar n$.

\begin{widetext}

\begin{figure}
	\includegraphics[width=1\linewidth]{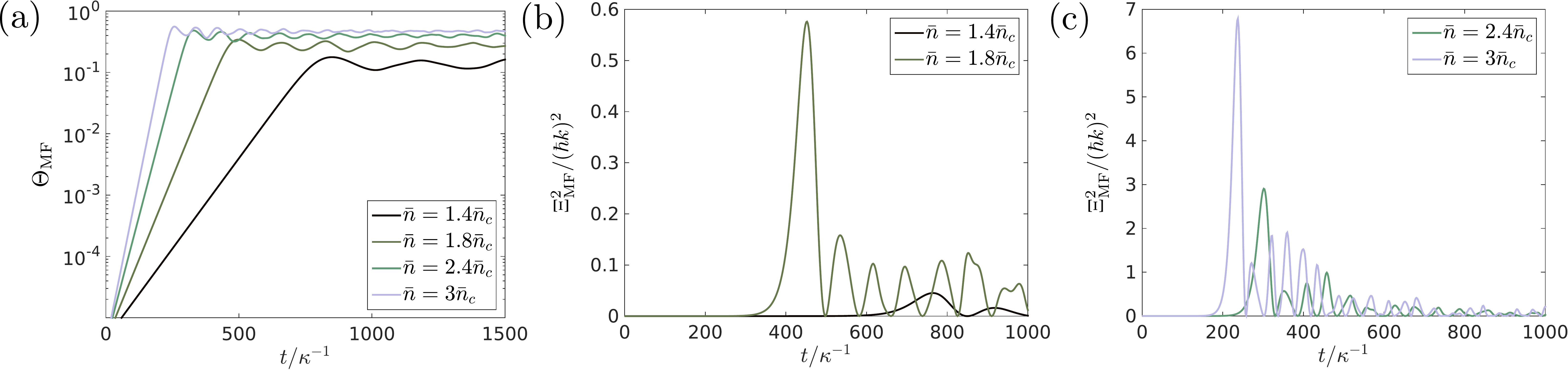}
\caption{(Color online) \label{vlasovplot} Time evolution of (a) the order parameter $\Theta_{\mathrm{MF}}$, Eq. \eqref{meanTheta},  and (b)-(c) parameter $\Xi_{\mathrm{MF}}^2$, Eq. \eqref{meanXi},  calculated by numerical integration of the Vlasov equation \eqref{vlasov} for different values of $\bar{n}$ and for $\Delta_c=-\kappa$. The initial distribution is given in Eq. \eqref{meanf0} with $\beta_0=2/(\hbar \kappa)$. %Since we see a linear behavior of $\Theta$ on a short timescale in (a) and the $y$-axis is given in a $\log$-scale be may expect a exponential increase of the order parameter. We also see that $\Theta$ relax with different exponents to different plateaus for the different $\bar{n}$. In (b) and (c) we see $\Xi^2$ for the same simulations. Whereby $\Xi^2$ seems to relax to 0 for all different $\bar{n}$, it is large at the violent relaxation. Furthermore the magnitude of $\Xi^2$ seems to increase rapidly with $\bar{n}$.
}
\end{figure}

We now analyse the initial exponential increase, which is in the regime where the Vlasov equation is a reliable approximation to the full dynamics, as we also verified in Ref. \cite{Schuetz:2015b}. In order to do so, we use a standard procedure, which is also detailed in Ref. \cite{Campa:2009,Griesser:2010}. For short times $t$ after the quench, we write the distribution as
\begin{align}
f_1(x,p,t)=f_1(p,0)+\delta f_1(x,p,t)\,,
\label{ansatzmeanf}
\end{align}
where $\delta f_1$ describes small fluctuations which can be due to the finite size of the system, and thus scale with $1/\sqrt{N}$. Using Eq. \eqref{ansatzmeanf} into the Vlasov equation \eqref{vlasov} and neglecting the terms of order $1/N$ we obtain the linearized Vlasov equation 
\begin{align}
 \frac{\partial\delta f_1}{\partial t}+\frac{p}{m}\frac{\partial \delta f_1}{\partial x}-\frac{\partial \delta V}{\partial x}\frac{\partial f_1(p,0)}{\partial p}&=0\label{vlasov linear}\,,
\end{align}
where $\delta V=V[\delta f_1(x,p,t)]$ and we dropped the argument of function $\delta f_1$. We seek for solutions of Eq. \eqref{vlasov linear} by means of the ansatz of Fourier waves with frequency $\omega$ and wave number $k$:
\begin{align}
\delta f_1&=g_1(p){\rm e}^{i(\omega t-kx)}+g_{-1}(p){\rm e}^{i(\omega t+kx)}, \label{deltaf}\,\\
\delta V&=2A\cos(kx){\rm e}^{i\omega t} \label{deltaV}\,,
\end{align} 
where $A\propto1/\sqrt{N}$ is some constant and the amplitudes $g_1(p)$ and $g_{-1}(p)$ are sole functions of the momentum $p$. The dispersion relation $\omega=\omega(k)$ can be derived after using Eqs. \eqref{deltaf} and \eqref{deltaV} in the linearized Vlasov equation, Eq. \eqref{vlasov linear}. By equating the coefficients of $\exp(ikx)$ and $\exp(-ikx)$ we get expressions for the functions $g_1(p)$ and $g_{-1}(p)$. With those expressions one finds the dispersion relation by using the definition $\delta V=V[\delta f_1]$ and Eqs. \eqref{meanpotentialnew} and \eqref{deltaV}: 
\begin{align}
\label{dispersionrelation}
0=1+\left(\hbar \Delta_c+i\frac{\hbar \kappa}{2}\hbar\omega\beta\right)\bar{n}\frac{1}{2}\int_{-\infty}^{\infty}dp\left(\frac{-k}{\frac{pk}{m}+\omega}+\frac{-k}{\frac{pk}{m}-\omega}\right)\partial_pf_1(p,0)\,.
\end{align}
This relation holds for any initial distribution that describes a uniform spatial density. We now use the Gaussian distribution in Eq. \eqref{meanf0} and obtain
\begin{align}
\label{newdispersionrelation}
0=1+\left(\hbar \Delta_c+i\frac{\hbar \kappa}{2}\hbar\omega\beta\right)\bar{n}\beta_0\left(1-\bar a\exp(-\bar a^2)\left(i\sqrt{\pi}-2\int_{0}^{\bar a}du\exp(u^2)\right)\right)\,,
\end{align}
where we defined $\bar a=\sqrt{\beta_0/(2m)} (m\omega/k)$. We then introduce $\bar b=i\bar a$ and $$\gamma=i\omega\,,$$ and cast Eq. \eqref{newdispersionrelation} into the form:
\begin{align}
\label{newdispersionrelation2}
0=1+\left(\hbar \Delta_c+\frac{\hbar \kappa}{2}\hbar\gamma\beta\right)\bar{n}\beta_0\left(1-\bar b\exp\left(\bar b^2\right)\left(\sqrt{\pi}-\int_{-\bar b}^{\bar b}du\exp(-u^2)\right)\right)\,,
\end{align}
\end{widetext}
where $\bar b\propto \gamma$. It can be shown that parameter $\gamma$, which solves Eq. \eqref{newdispersionrelation2}, is a real number. Therefore, $\omega$ is an imaginary number. In particular, if $\gamma<0$ both Eqs. \eqref{deltaf} and \eqref{deltaV} describe fluctuations which are exponentially damped and therefore $f_1(x,p,t)$ will tend to the initial distribution, which is stable. If instead the solution of Eq. \eqref{newdispersionrelation} gives $\gamma>0$, the initial distribution is unstable against fluctuations. The value $\gamma=0$ separates the two regimes. After setting $\gamma=0$ in Eq. \eqref{newdispersionrelation2} we thus get the critical condition
\begin{align}
1=-\hbar \Delta_c\bar{n}\beta_0\,, \label{criticalconditionvlasov}
\end{align}
which connects $\Delta_c$, $\bar n$, and the initial temperature $1/\beta_0$, which is an external parameter. If $\beta_0$ coincides with the value in Eq. \eqref{finaltemperature}, then Eq. \eqref{criticalconditionvlasov} corresponds to the same relation  as in Eq. \eqref{threshold}, which defines the critical value of $\bar n$ for self-organization.  For the values of the parameters, for which $\gamma>0$, the uniform distribution is unstable and tends to form a grating at the wave vector $k$ of the resonator with exponential increase, giving rise to a violent relaxation. Parameter  $\gamma$ gives the rate at which the amplitude of this density modulation grows. 

\begin{figure}
	\includegraphics[width=0.8\linewidth]{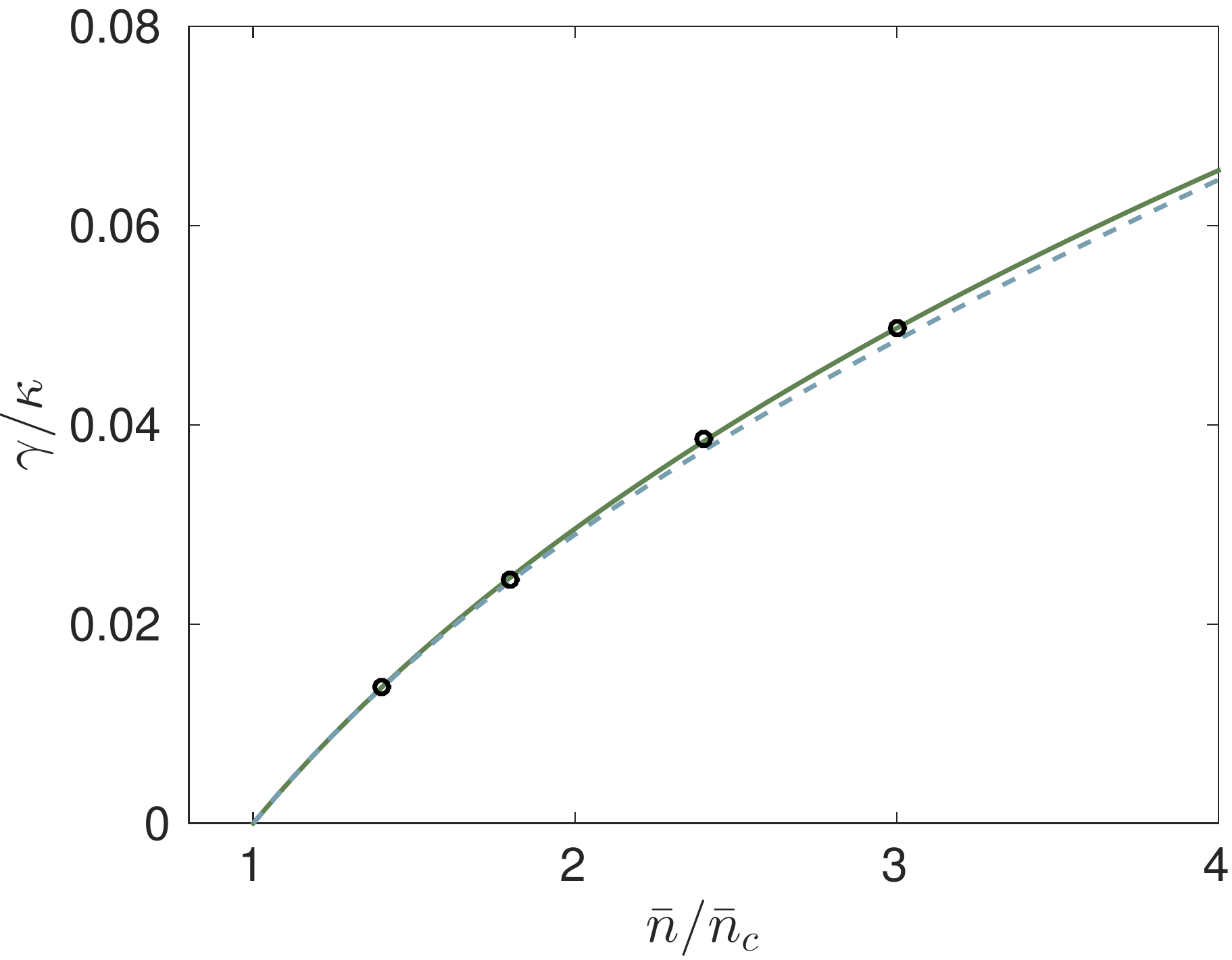}
	\caption{(Color online) \label{exponentplot} Slope $\gamma$ of the initial increase of $\Theta_{\mathrm{MF}}$. The dots are extracted by fitting the curve obtained from the numerical simulations in Fig. \ref{vlasovplot}, the dashed line is the value predicted by Eq. \eqref{Simon:freund}, which well agrees with Eq. \eqref{newdispersionrelation2} (solid line). For these parameters the threshold for the Vlasov stability, Eq. \eqref{criticalconditionvlasov}, reads $1=\bar{n}/\bar{n}_c$.}
%	
%	The dashed line is an approximate solution that has to be derived self-consistently in the lime $|\Delta_c|\gg|\gamma|$. It reads $\gamma=\frac{\sigma k}{m}\left(1-p\chi\right)\frac{-\ln\left(\frac{\chi}{1.135}\right)+\ln\left(1-p\chi\right)}{1.4\left(1-p\chi\right)+\frac{\hbar k}{2m}\sigma\kappa\beta/|\Delta_c|}$, with $\chi=-\hbar|\Delta_c|\bar{n}\beta_0=(\bar n/\bar n_c)(\beta_0/\beta)$, $\sigma=\sqrt{m/\beta_0}$ and $p=27/227$. 
%	
%	The solid line represents the solution $\gamma$ of equation \eqref{newdispersionrelation2} for different $\bar {n}$. For this numeric calculation we put $\Delta_c=-\kappa$ and $\beta_0=2/(\hbar \kappa)$. For these parameters the threshold for the Vlasov stability \eqref{criticalconditionvlasov} reads $1=\bar{n}/\bar{n}_c$. The dashed line is an approximate solution that has to be derived self-consistently in the lime $|\Delta_c|\gg|\gamma|$. It reads $\gamma=\frac{\sigma k}{m}\left(1-p\chi\right)\frac{-\ln\left(\frac{\chi}{1.135}\right)+\ln\left(1-p\chi\right)}{1.4\left(1-p\chi\right)+\frac{\hbar k}{2m}\sigma\kappa\beta/|\Delta_c|}$, with $\chi=-\hbar|\Delta_c|\bar{n}\beta_0=(\bar n/\bar n_c)(\beta_0/\beta)$, $\sigma=\sqrt{m/\beta_0}$ and $p=27/227$. The dots represent the slope of the linear behavior of the four trajectories shown in Fig. \ref{vlasovplot} (a). We see that the four dots are in very good agreement with the predicted solid line. This shows that the linearized Vlasov equation can explain the short time behavior very well. The dashed line fits to the dots and the solid curve at least for small $\bar{n}/\bar{n}_c$.}
\end{figure}

Figure \ref{exponentplot} compares the value of $\gamma$ extracted by fitting the exponential increase of $\Theta_{\mathrm{MF}}$ in the first phase of the dynamics of Fig. \ref{vlasovplot} and for different values of $\bar n$, with the one determined by Eq. \eqref{newdispersionrelation2}, showing very good agreement. In particular, we note that in the limit $|\Delta_c|\gg|\gamma|$ Eq. \eqref{newdispersionrelation2} can be reduced to the form \cite{Karagiannidis:2007}
\begin{equation}
\gamma=\omega_0\left(1-p\chi\right)\frac{\ln\left(\frac{\chi}{1.135}\right)-\ln\left(1-p\chi\right)}{1.4\left(1-p\chi\right)+\hbar\kappa\beta\omega_0/(2|\Delta_c|)}\,,
\label{Simon:freund} 
\end{equation}
with $\chi=\hbar|\Delta_c|\bar{n}\beta_0=(\bar n/\bar n_c)(\beta_0/\beta)$, $\omega_0=\sqrt{2\omega_r/(\hbar\beta_0)}$ and $p=27/227$.

%An approximate form of Eq. \eqref{newdispersionrelation2} can be found for $|\Delta_c|\gg|\gamma|$, which delivers the value of $\gamma$ as a function of the other physical parameters, and in particular $\chi=-\hbar|\Delta_c|\bar{n}\beta_0=(\bar n/\bar n_c)(\beta_0/\beta)$:
%\begin{align}
%\gamma=\frac{\sigma k}{m}\left(1-p\chi\right)\frac{-\lpg\left(\frac{\chi}{1.135}\right)+\ln\left(1-p\chi\right)}{1.4\left(1-p\chi\right)+\frac{\hbar k}{2m}\sigma\kappa\beta/|\Delta_c|}.\label{approximate solution}
%\end{align}
%with $p=0.135/1.135$.
%
%\begin{align}
%\frac{\gamma}{-\Delta_c}=-\frac{\ln\left(\frac{1.135}{-\hbar\Delta_c\bar{n}\beta_0}-0.135\right)}{-1,4\frac{m\Delta_c}{\sigma k}+\frac{\frac{\hbar\kappa}{2}\beta}{1+\frac{0.135}{1.135}\hbar \Delta_c\bar{n}\beta_0}}.\label{approximate solution}
%\end{align}

\section{Validity of the mean-field ansatz}
\label{Sec:Limits}

The mean-field treatment is based on the assumption that the distribution function for the $N$ particle can be approximated by the product of the single-particle distribution. This ansatz thus discards interparticle correlations which emerge from the photon-mediated interactions: the factorized ansatz is very different from the form of the distributions one obtains from the full $N$-particle FPE \cite{Schuetz:2014,Schuetz:2015}. Nevertheless, the assumption still captures essential features of the short-time dynamics of distributions, which have initially the form of Eq. \eqref{factorization}. We will follow the procedure illustrated in Ref. \cite{Campa:2009, NardiniPHD} and study the validity of the mean-field ansatz within a BBGKY  hierarchy, which we derive from the $N$-particle FPE, Eq. \eqref{FPE}. We will particularly focus on the dynamics of two-particle correlations and determine the characteristic time scale of their dynamics. \\
For convenience, we introduce the vectors ${\bf x}=(x_1,...,x_N)^T$ and ${\bf p}=(p_1,...,p_N)^T$, and define $f_N({\bf x};{\bf p};t)\equiv f_N(x_1,...,x_N;p_1,...,p_N;t)$.

%In the last section we derived some properties of the stationary state. This stationary state can be described in the mean-field where we are just integrating out the momentum and position of all atoms but one in the full $N$-particle description. Unfortunately one can not describe the $N$-particle dynamics towards the equilibrium because the mean-field ansatz \eqref{factorization} is only valid for short timescales. In this section we want to justify this approach. First we will give a general treatment of the FPE in terms of a BBGKY hierarchy. After that we will look at the dynamics of two-particle correlations and show that these correlations evolve on a larger timescale. This will be shown by deriving the Lenard-Balescu equation.
%
\subsection{BBGKY hierarchy of the photon-mediated Fokker-Planck equation}
\label{Sec:BBGKY}
For the derivation of the BBGKY hierarchy we assume that the energy of the system is finite. This corresponds to assume that the limit holds:
\begin{align}\label{vanishesinfinity}
\lim\limits_{|\mathbf{p}|\to\infty}f_N(\mathbf{x};\mathbf{p};t)=0,
\end{align}
where $|{\bf p}|=\sqrt{\sum_{i=1}^Np_i^2}$, and that expectation values of all moments exist.
Furthermore $f_N$ is periodic with wavelength $\lambda$  in every  $x_i$, which implies
\begin{align}\label{periodic}
f_N(\mathbf{x}+\lambda \mathbf{ z};\mathbf {p};t)=f_N(\mathbf{x};\mathbf {p};t),
\end{align}
for every ${\mathbf z}\in\mathbb{Z}^N$.
Distribution function $f_N$ is invariant under particle exchange, which we can express by means of the permutation matrix ${\bf P}$, such that:
\begin{align}\label{permutation}
f_N(\mathbf{P}\mathbf{x};\mathbf{P}\mathbf{p};t)=f_N(\mathbf{x};\mathbf{p};t),
\end{align}
where each row and column of ${\bf P}$ contain only one entry different from zero and equal to 1.\\
In order to derive the BBGKY hierarchy of the FPE in Eq. \eqref{FPE} we first define the $l$-particle distribution function:
\begin{align}
f_l=\int_{0}^{\lambda}\frac{{\rm d}x_{l+1}}{\lambda}\int_{-\infty}^{\infty}{\rm d}p_{l+1}\ldots\int_{0}^{\lambda}\frac{{\rm d}x_{N}}{\lambda}\int_{-\infty}^{\infty}{\rm d}p_Nf_N\label{fldef}\,,
\end{align}
where $f_l$ inherits the three properties in Eqs. \eqref{vanishesinfinity}, \eqref{periodic} and \eqref{permutation} from $f_N$.
Index $l$ takes the value $l=1,\ldots,N$, such that for $l=1$ the distribution $f_l$ is the single-particle phase-space function, and for $l=N$ it describes the $N$ particle state. The evolution of $f_l$ is found from Eq. \eqref{FPE} after integrating out the other $N-l$ particle variables, and can be cast in the form
\begin{align}\label{fl}
\frac{\partial f_l}{\partial t}&=\sum_{j=1}^{l}\left(\mathcal L_j^{(l)}f_l+\mathcal G_j^{(l)}[f_{l+1}]\right)\,,
\end{align}
where the first operator on the RHS solely depends on the variables of the $l$ particles and reads
\begin{eqnarray}
&&\mathcal L_j^{(l)}f_l= -\frac{\partial }{\partial x_j}\frac{p_j}{m}f_l\label{L:l}\\
&&-S^2 \frac{\partial}{\partial p_j}\sum_{i=1}^{l}\left(F_0\cos(kx_i)+\Gamma_0\sin(kx_i)p_i\right)\sin(kx_j)f_l\notag\\
&&+S^2\frac{\partial}{\partial p_j}\sum_{i=1}^{l}\left(D_0\frac{\partial}{\partial p_i} +\eta_0\frac{\partial}{\partial x_i}\right)\sin(kx_i)\sin(kx_j)f_l\notag\,.
%&&+\sum_{i=1}^{l}\frac{\partial^2}{\partial p_j\partial x_i} \eta_0S^2 \sin(kx_i)\sin(kx_j)f_l,\label{L:l}
\end{eqnarray}
The second operator, instead, depends nonlinearly on the $(l+1)$-particle distribution function. This term vanishes when $l=N$, while for $l<N$ it describes the dynamics of correlations, which are established by the interparticle potential.  It reads
\begin{align}
\mathcal G_j^{(l)}[f_{l+1}]= -S^2(N&-l)\frac{\partial}{\partial p_j}\sin(kx_j)\nonumber\\
&\times\left(F_0\Theta_l[f_{l+1}]+\Gamma_0\Xi_l[f_{l+1}]\right)\,,
\label{G:l}
\end{align}
where
\begin{align}
\Theta_l[f_{l+1}]&= \int_{0}^{\lambda}\frac{{\rm d}x_{l+1}}{\lambda}\int_{-\infty}^{\infty}{\rm d}p_{l+1}\cos(kx_{l+1})f_{l+1}\label{Theta(f)neu}\\
\Xi_l[f_{l+1}]&=\int_{0}^{\lambda}\frac{{\rm d}x_{l+1}}{\lambda}\int_{-\infty}^{\infty}{\rm d}p_{l+1}\sin(kx_{l+1})p_{l+1}f_{l+1}\label{Xi(f)neu}\,,
\end{align}
while $\Theta_0[f_1]=\Theta_{\mathrm{MF}}$ and $\Xi_0[f_1]=\Xi_{\mathrm{MF}}$. Note that when the factorization ansatz of Eq. \eqref{factorization} holds, then $\Theta_1[f_2]=\Theta_{\mathrm{MF}}f_1$ and $\Xi_1[f_2]=\Xi_{\mathrm{MF}}f_1$.  A closed set of equations for $f_l$ can be thus strictly obtained for $l=N$, giving Eq. \eqref{FPE}, or for $S=0$, hence in absence of the cavity field. 

\subsection{The Lenard-Balescu equation}
\label{Sec:LenardBalescu}

\begin{figure}
	\includegraphics[width=0.8\linewidth]{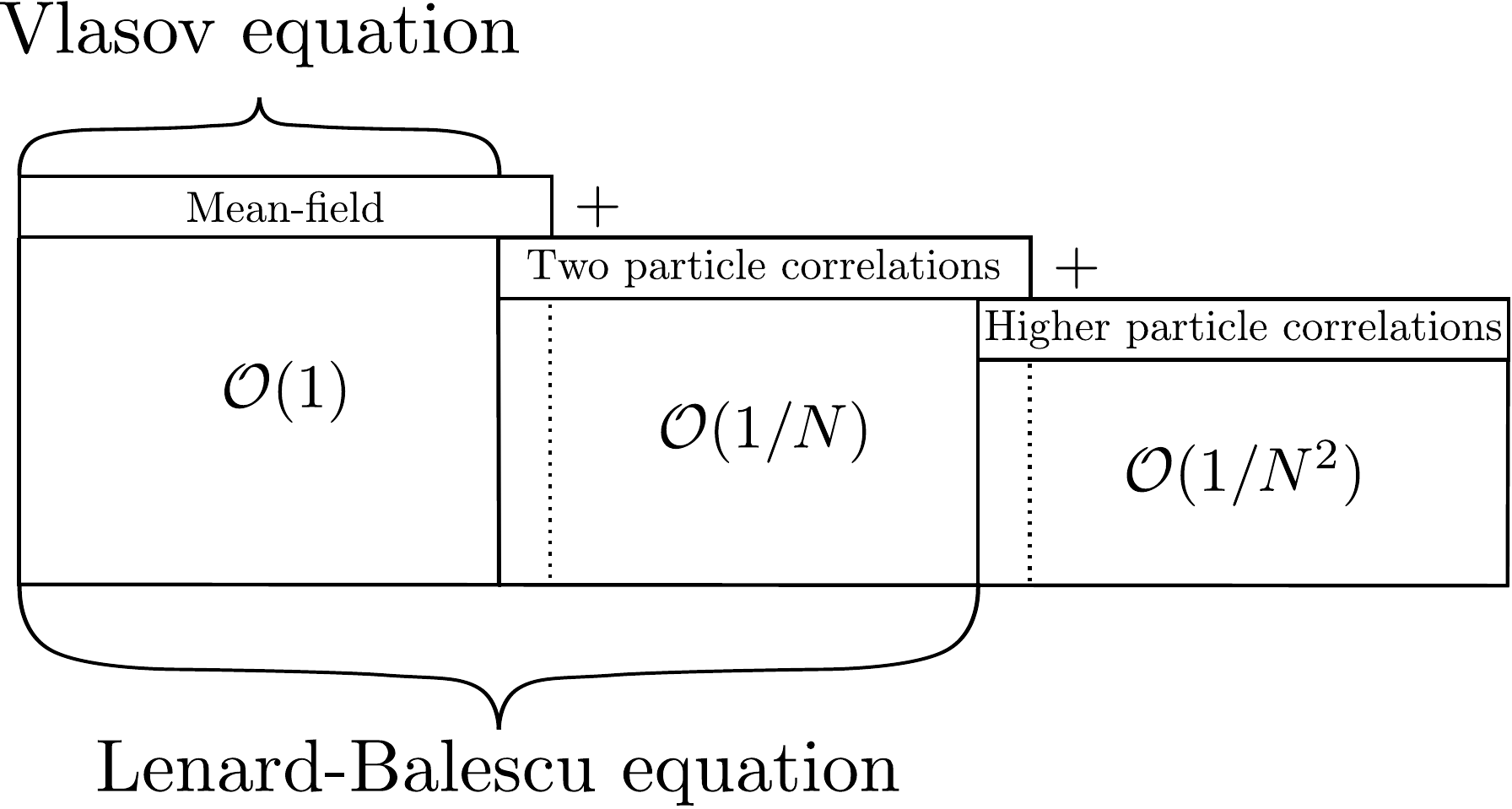}
	\caption{\label{fig:BBGKY}Illustration of the order of magnitude of the corrections of the Vlasov and of the Lenard-Balescu equations, and of which type of correlations they include.}
\end{figure}

For $l=2$ we can generally decompose the distribution function into two terms:
\begin{align}
f_2(x_1,x_2,p_1,p_2)&=f_1(x_1,p_1)f_1(x_2,p_2)+g_2(x_1,x_2,p_1,p_2)\label{g2}\,,
\end{align}
where the first term on the RHS is the mean-field term and the second term describes all corrections beyond mean field. When at $t=0$ the distribution function is factorized in a form like Eq. \eqref{factorization}, the dynamics beyond mean field will tend to build correlations which are described by $g_2$.  We obtain the mean-field FPE, Eq. \eqref{meanFPE} by performing the approximation  $\mathcal G_1^{(1)}[f_{2}]\to\mathcal G_1^{(0)}[f_{1}]f_1$. In the following we analyse the regime in which this approximation is justified by studying the equation describing the evolution of the function $g_2$ under some approximation, which permits us to truncate the BBGKY hierarchy till second order. This equation is known in the literature as Lenard-Balescu equation \cite{Campa:2009}, and it will allow us to identify a time-scale where the mean-field treatment provides reliable predictions. 
\begin{widetext}
In order to derive the Lenard-Balescu equation we first consider the distribution function for $l=3$. Using the same type of decomposition as in Eq. \eqref{g2}, this can be written as
\begin{align*}
f_3(x_1,x_2,x_3,p_1,p_2,p_3)&=f_1(x_1,p_1)f_1(x_2,p_2)f_1(x_3,p_3)\\
&+\sum_{i,j,k=1}^3|\epsilon_{ijk}|f_1(x_i,p_i)g_2(x_j,p_j,x_k,p_k)\\
&+g_3(x_1,x_2,x_3,p_1,p_2,p_3),
\end{align*}
where $\epsilon_{ijk}$ is the Levi-Civita tensor and $g_3$ describes all three-body correlations which cannot be written as a function of $f_1$ and/or $f_2$. 
We assume now that $g_3$ is of higher order (from the treatment below we will see that $g_3\propto 1/N^2$) and drop $g_3$ in the equation describing the dynamics of $f_2$, Eq. \eqref{fl}. By means of this assumption we obtain two coupled equations for $f_1$ and $f_2$, which can be then cast into the Lenard-Balescu equations for $f_1$ and $g_2$ using Eq. \eqref{g2} and which read 
\begin{subequations}
\begin{align}
\frac{\partial f_1}{\partial t}=&\mathcal L^{(1)}f_1+\mathcal G^{(1)}[f_1]f_1+\mathcal G^{(1)}[g_2]\label{Balescu1}\\
\frac{\partial g_2}{\partial t}= &-\frac{\partial }{\partial x_1}\frac{p_2}{m}g_2-\frac{\partial }{\partial x_2}\frac{p_1}{m}g_2\label{Balescu2}\\
&-S^2\sum_{j=1}^2\sum_{i\neq j}\frac{\partial}{\partial p_j}F_0\sin(kx_j)\left(\cos(kx_i)-\Theta_{\mathrm{MF}}[f_{1}]\right)f_{1}f_{1}\notag\\
&-S^2\sum_{j=1}^2\sum_{i\neq j}\frac{\partial}{\partial p_j}\Gamma_0\sin(kx_j)\left(\sin(kx_i)p_i-\Xi_{\mathrm{MF}}[f_{1}]\right)f_{1}f_{1}\notag\\
&+S^2\sum_{j=1}^2\sum_{i\neq j}\frac{\partial}{\partial p_j} \sin(kx_j)\left(D_0\frac{\partial}{\partial p_i}+\eta_0\frac{\partial}{\partial x_i}\right)\sin(kx_i)f_1f_1\notag\\
&-NS^2F_0\sum_{j=1}^2\sum_{i\neq j}\frac{\partial}{\partial p_j}\sin(kx_j)\left(\Theta_1[g_{2}]_if_1(x_j,p_j)+\Theta_{\mathrm{MF}}[f_{1}]g_2\right)\notag\\
&-NS^2\Gamma_0\sum_{j=1}^2\sum_{i\neq j}\frac{\partial}{\partial p_j}\sin(kx_j)\left(\Xi_1[g_{2}]_if_1(x_j,p_j)+\Xi_{\mathrm{MF}}[f_{1}]g_2\right)\,,\notag
\end{align}
\label{LenardBalescu}
\end{subequations}
where we specified the arguments when necessary, and introduced the notation $\Theta_1[g_{2}]_i$ and $\Xi_1[g_{2}]_i$ to indicate that these are functions of $(x_i,p_i)$. 

The validity of the mean-field FPE, Eq. \eqref{meanFPE}, relies on whether one can discard term $\mathcal G^{(1)}[g_2]$ in the RHS of Eq. \eqref{Balescu1}. Let us recall the thermodynamic limit for which $S^2\sim 1/N$. If we now assume that $g_2$ is of order $1/N$ with respect to $f_1$, then the term $\mathcal G^{(1)}[g_2]$ is of order $1/N$ with respect to $\mathcal G^{(1)}[f_1]f_1$. A detailed analysis of Eq. \eqref{Balescu2} shows that, if $g_2\sim 1/N$ at $t=0$, this scaling is preserved by the dynamics. In fact, (i) the first line on the RHS of Eq. \eqref{Balescu2} gives a scaling with $1/N$ because it is proportional to $g_2$, while all other quantities are independent of $N$, (ii) the second, third, and fourth lines are all proportional to $S^2\sim 1/N$, (iii) the last two lines scale with $NS^2 g_2\sim 1/N$. Therefore, for sufficiently short times the contribution of $g_2$ to the dynamics in the mean-field equation can be neglected. 

We note that in Eq. \eqref{Balescu1} the term $\mathcal L^{(1)}f_1$ has also components which scale with $1/N$. If one consistently neglects all terms scaling with $1/N$, then Eq. \eqref{Balescu1} reduces to the Vlasov equation, Eq. \eqref{vlasov}, and therefore also neglects the diffusion processes leading to equilibrium. Figure \ref{fig:BBGKY} illustrates the order of magnitude of the corrections  to the Vlasov and Lenard-Balescu equations, as well as the type of correlations that these describe.
\end{widetext}

	\begin{figure}
		\includegraphics[width=1\linewidth]{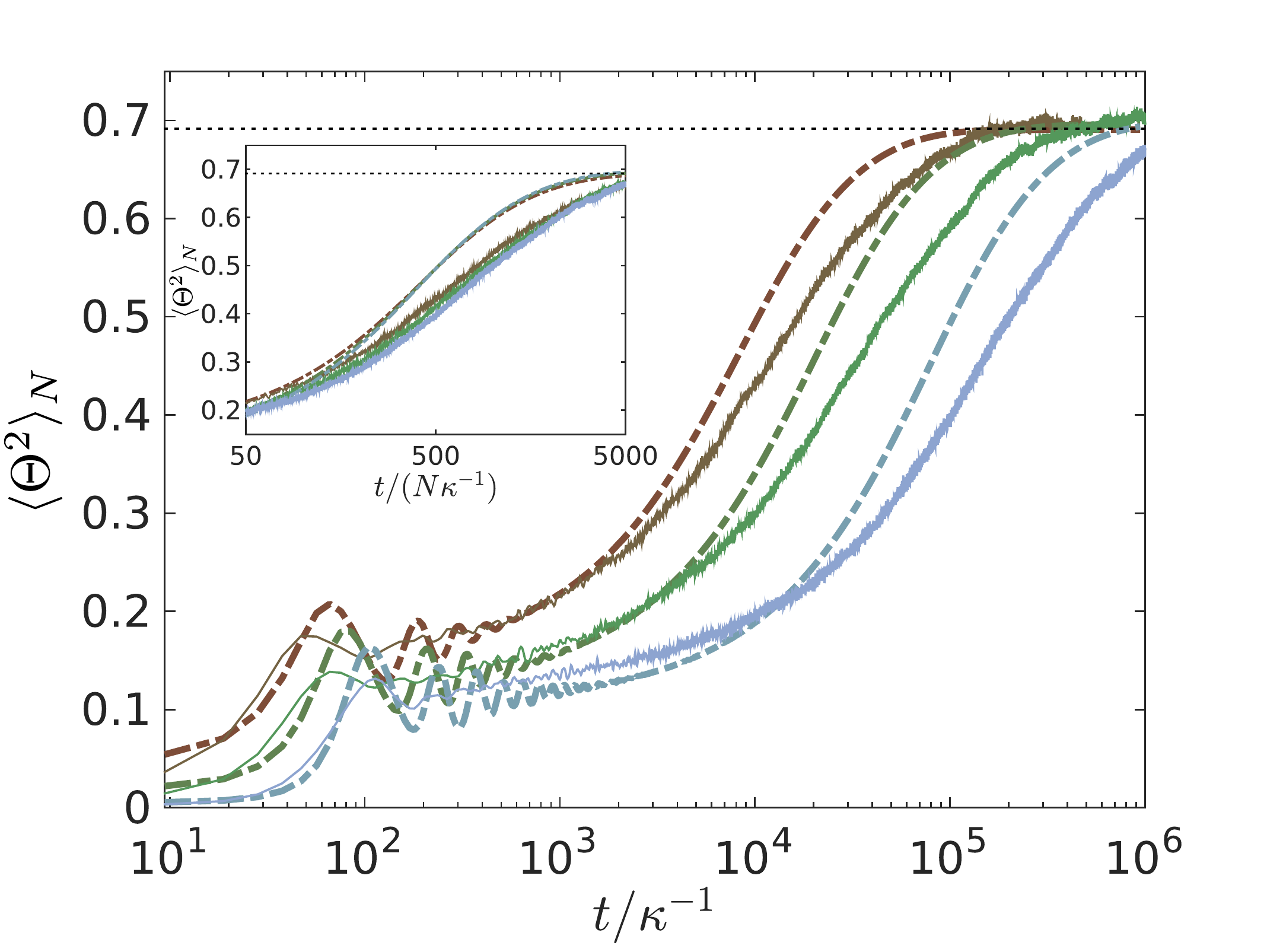}
\caption{(Color online) \label{Dynamiccomparison} Time evolution of the squared order parameter evaluated by numerically simulating (dashed-dotted lines)  the mean-field FPE, Eq. \eqref{meanFPE} and (solid lines) the $N$-particle FPE of Eq. \eqref{FPE}. The curves correspond to different particle numbers $N=20$ (brown), $N=50$ (green) and $N=200$ (blue) and are calculated taking $\Delta_c=-\kappa$ and $\bar{n}=2\bar{n}_c$. The number $\mathcal T$ of trajectories taken for the $N$-body FPE is $\mathcal T=1000$ for $N=20$, $\mathcal T=500$ for $N=50$, $\mathcal T=100$ for $N=200$ (see Ref. \cite{Schuetz:2013,Schuetz:2015} for details on the simulations). The horizontal dotted line indicates the asymptotic value of the squared order parameter. The inset shows the curves of the onset with the time axis rescaled by $N$. Note that the initial distribution of the full $N$-body FPE is the one which statistically corresponds to a spatially uniform distribution with the same temperature as the asymptotic one. Therefore, the value of $\langle\Theta^2\rangle_N$ at $t=0$ does not vanish due to finite size effects. In order to compare these dynamics with the mean-field FPE, we have taken into account these finite-size effects in the initial mean-field distribution given by $\tilde{f}_0(x,p)=(1+\delta_N\cos(kx))f_1(p,0)$ where $f_1$ is given in Eq. \eqref{meanf0} and $\delta_N$ is a spatial modulation depending on $N$.}
	\end{figure}
\subsection{Mean-Field versus full $N$-atom dynamics}

In order to complete our analysis of the limits of validity of the mean-field ansatz, we now compare its predictions with the ones obtained by numerical simulations of the $N$-particle FPE of Eq. \eqref{FPE}. The latter are performed by means of stochastic differential equations (see Refs. \cite{Schuetz:2013, Schuetz:2015} for details). We focus now on the evolution of the expectation value of $\Theta^2$, which explicitly depends on two-particle correlations and scales the strength of the conservative many-body potential. We recall the definition $\langle .\rangle_N$ in order to indicate the mean value of a $N$-particle observable taken over the $N$-particle distribution $f_N$.\\ 
Figure \ref{Dynamiccomparison} compares the $N$-particle description where the evolution of $f_N$ is governed by FPE \eqref{FPE} (solid line) and the mean-field description, where $f_N({\bf x};{\bf p};t)=f_1(x_1,p_1;t)...f_1(x_N,p_N;t)$ and the evolution of $f_1$ is governed by the mean-field FPE \eqref{meanFPE} (dashed-dotted line).
The curves are plotted as a function of time and for different particle numbers, $N=20,50,200$, where the parameter $S^2$ has been rescaled according to our thermodynamic  limit so to warrant a threshold $\bar n_c$ which is independent on $N$. The parameters have been fixed so that initially the distribution is spatially uniform,  while the momentum distribution is a Gaussian whose width coincides with the asymptotic temperature of the dynamics, Eq. \eqref{finaltemperature}. The strength of the field is such that $\bar n=2\bar n_c$, therefore the asymptotic spatial distribution is a Bragg grating with $|\Theta_{\mathrm{MF}}|\sim 0.83$. The dynamics we observe is the one which leads to the formation of the Bragg gratings starting from a uniform spatial distribution, and exhibit three stages, which have been extensively discussed in Ref. \cite{Schuetz:2015b}: a violent relaxation,  a prethermalized phase, and a slow approach to equilibrium. The full lines are the simulation of the full FPE, the dashed lines the corresponding mean-field prediction, which indeed qualitatively reproduces the three-stage dynamics. 

The violent relaxation is a stage of the dynamics where there is a good agreement between mean-field and $N$-body FPE. This is the short-time regime where the Vlasov equation, Eq. \eqref{vlasov}, is valid, and the behaviour of the $N$-body FPE is reproduced by the one observed numerically integrating the Vlasov equation, see Fig. \ref{vlasovplot}(a). This has been also verified in Ref. \cite{Schuetz:2015b}. The prethermalized regime is also predicted by the Vlasov equation, see Fig. \ref{vlasovplot}(a). The mean-field FPE, however, provides a more accurate description and qualitatively reproduces the $N$-body FPE. Nevertheless, a clear difference between mean-field and $N$-body dynamics is found at the onset of the prethermalized stage: In fact, the oscillations are damped at a faster rate in the $N$-body FPE. Apart from this difference, there is a qualitative agreement between mean-field and $N$-body FPE also for this stage. 

While both mean-field and $N$-body FPE agree in the asymptotic value, we observe a striking difference between the two results in the relaxation to equilibrium after prethermalization. This is the stage where the role of dissipation and diffusion becomes relevant, as shown in Ref. \cite{Schuetz:2015b} by comparing this behaviour with the one, where the dynamics is only due to the Hamiltonian term. In particular,  the relaxation time scale predicted by the full simulation is about one-order of magnitude longer than the corresponding mean-field prediction. This becomes even more evident by plotting the curves rescaling the time axis with $N$, as visible in the inset. The curves of the mean-field FPE collapse to one curve, whereby the ones of the $N$-body FPE collapse to a significantly different curve.
% The linear behaviour thus suggests that the approach to equilibrium occurs with an exponential damping, whose rate scales with $1/N^{\delta}$. The exponent $\delta$ of the full FPE thus seems to be larger than the one predicted by the mean-field FPE.%, being the corresponding slope smaller. % and suggests a superlinear scaling of relaxation, typical of quasi-stationary states of long-range interacting systems \cite{HMF,Campa:2009}. While the quasi-stationary states analysed so far were due to Hamiltonian dynamics, here they are instead the result of the interplay between conservative and dissipative long-range forces.

%We note that the mean-field FPE of Eq. \eqref{meanFPE} shares several analogies with the Brownian Mean-field Model of Ref. \cite{Chavanis:2014}, where damping and diffusion coefficients do not have a spatial off-diagonal structure. In particular, we attribute the difference between the mean-field FPE and the $N$-body FPE to the off-diagonal terms of the incoherent part of the dynamics, which is averaged out in the factorization ansatz of mean-field. 
%In order to verify this hypothesis, we perform simulations for the same parameters with the $N$-body FPE, where we set to zero all off-diagonal elements of friction and diffusion. Hence, in lines \eqref{friction} and \eqref{diffusion} we set to zero the terms with $i\neq j$. The results of these simulations are shown in Fig. \ref{Dynamiccomparison}(b) and are in very good agreement with the curves of the mean-field treatment for all times. 

Let us now summarize these results. First, the short time behaviour of the fluctuations of the order parameter are well described by the mean-field equation, and in particular by the Vlasov equation. This is well understood in terms of the typical contributions to the dynamics: For short times the dominant contributions are indeed the terms of Eq. \eqref{vlasov} and interparticle correlations are small, as we argued in the previous section. Discrepancies are due to finite size effects. The prethermalized regime, moreover, exhibits a good agreement between mean-field and full dynamics. This regime is dominated by the Hamiltonian dynamics, and the results show that Hamiltonian dynamics with long-range interactions is well reproduced by the mean-field description. %Hamiltonian dynamics have been so far identified to be responsible for quasi-stationary states, and these have been understood in terms of Vlasov-stable solutions. The presence of noise and dissipation, instead, has been considered detrimental, turning the superlinear scaling with $N$ of the time scale leading to equilibrium to a linear scaling \cite{Gupta:2010,Chavanis:2014}. Remarkably, it is here instead noise and diffusion that induce such scaling. This occurs on long time scale, and this peculiar feature is lost in the mean-field treatment, since the long-range features of dissipation and diffusion are averaged out by means of the factorization ansatz.   
Big deviations instead appear for long times, where the mean field ansatz is expected to fail and at the time scales dominated by relaxation to the stationary state.

\FloatBarrier

\section{Conclusions}
\label{Sec:Conclusions}

In this work we have systematically developed a mean-field description of the self-organization dynamics of atoms in a high-finesse cavity. The predictions of the mean-field model have been explored at equilibrium and out-of-equilibrium, its limits of validity have been tested by comparing them with the ones of the $N$-body FPE. We have found that the mean-field equation provides an excellent description of the dynamics  when this is prevailingly Hamiltonian. It further describes the equilibrium properties of single-particle observables, including the asymptotic temperature and the order parameter. It fails, however, to reproduce the long-time out-of-equilibrium dynamics. 

%This discrepancy is instructive and peculiar. It shows that in this system the superlinear slowing down of relaxation induced by the off-diagonal noise leads to the existence of quasi-stationary states, which cannot be described by means of a mean-field model. This is quite different from the type of dynamics of the Hamiltonian-mean-field model, where quasi-stationary states result from the Hamiltonian dynamics and where the typical superlinear scaling with $N$ is suppressed by uncorrelated noise \cite{HMF,Campa:2009,Chavanis:2014}. Here, instead, noise exhibits long-range spatial correlations and is responsible for the appearance of superlinear scaling. 

Despite these differences, this analysis shows that from the mean-field model one can analytically extract several predictions on the system dynamics. It is indeed remarkable that several predictions reproduce in the corresponding limits the ones obtained by means of other theoretical treatments, some of which start from a fully quantum mechanical treatment for the atoms. This on the one hand leads us to conjecture that quantum fluctuations play a marginal role in determining the steady state properties of the cavity field. It further urges one to develop a full quantum kinetic theory, analogous to the full $N$-body semiclassical theory, which shall overcome all limitations of simplifying theoretical assumptions performed so far. Only such a model, in fact, can give full access to the dynamical interplay between matter waves and cavity photons. 

\section*{Acknowledgments}
The authors are grateful to G. Manfredi and C. Nardini for insightful discussions. This work was supported by the German Research Foundation (DACH project: "Quantum crystals of matter and light").

\appendix
\section{Cavity field correlation function at steady state} \label{AppendixA}

Experimentally accessible quantities are the correlation functions of the field at the cavity output, which allows one to monitor the atoms state and is proportional to the intracavity field. In our formalism, the intracavity field is closely connected to the atomic state by the relation $E_{\rm cav}\propto \sqrt{N\bar n}\Theta$, therefore the correlation functions of the cavity field are proportional to the correlation functions of the magnetization $\Theta$ \cite{Schuetz:2013,Schuetz:2015}. In the following we determine the autocorrelation function of the magnetization, which can be detected by means of the first-order correlation function of the field, and the fourth-moment of the magnetization $\langle\Theta^4\rangle_{N}$. As we showed in Ref. \cite{Schuetz:2015}, in fact, $\langle\Theta^4\rangle_{N}$ delivers the value of the intensity-intensity correlation of the field at zero-time delay and at zero order in the retardation effects. 

\subsection{Field intensity across the transition}

We first determine the intracavity photon number $n_{\rm cav}$ at steady state for $\bar n$ below, at, and above threshold. For this purpose we use the relation \cite{Schuetz:2013,Schuetz:2015}
\begin{equation}
\label{n:cav}
n_{\rm cav}=N\bar{n}\langle \Theta^2\rangle_{N}\,,
\end{equation}
which, by introducing $\alpha=\bar n/\bar n_c$, can be cast in the form (see also the Appendix \ref{fluxexponent})
\begin{eqnarray}
n_{\rm cav}=\frac{1}{2}\bar{n}_c+\bar n\frac{\partial}{\partial\alpha}\mathcal G(\alpha)\,,
\label{To:show}
\end{eqnarray}
where
\begin{equation}
\label{G:alpha}
\mathcal G(\alpha)=\ln\left( \int_{-\infty}^{\infty} dy\exp\left[-N\left(\alpha y^2-\ln(I_0(2\alpha y))\right)\right]\right)\,.
\end{equation}
We then analyse the prediction of this expression close to threshold, for $\bar n\sim \bar n_c$ and thus $\alpha\sim 1$. For this purpose we expand the exponent of $\mathcal G(\alpha)$ about the value $y=0$ and consider the behaviour of $n_{\rm cav}$ for $\alpha\to 1^-$, hence for $\bar n<\bar n_c$ but sufficiently close to the transition point, so that the truncation of the expansion is valid. In this limit we find 
\begin{equation}
\label{n:cav:below}
n_{\rm cav}\approx \frac{\bar{n}_c^2/2}{\bar{n}_c-\bar{n}}\,,
\end{equation}
where the details of the derivation are reported in the Appendix \ref{fluxexponent}. The value at the transition point is reported at leading order in $N$ and reads (see Appendix \ref{fluxexponent}):
\begin{align}
\label{n:cav:at}
n_{\rm cav}\approx2\sqrt{N}\bar{n}_c\frac{\Gamma\left[\frac{3}{4}\right]}{\Gamma\left[\frac{1}{4}\right]}\,,
\end{align}
where $\Gamma[x]$ denotes the Gamma function \cite{Abramowitz}. 

\begin{figure}
	\includegraphics[width=1\linewidth]{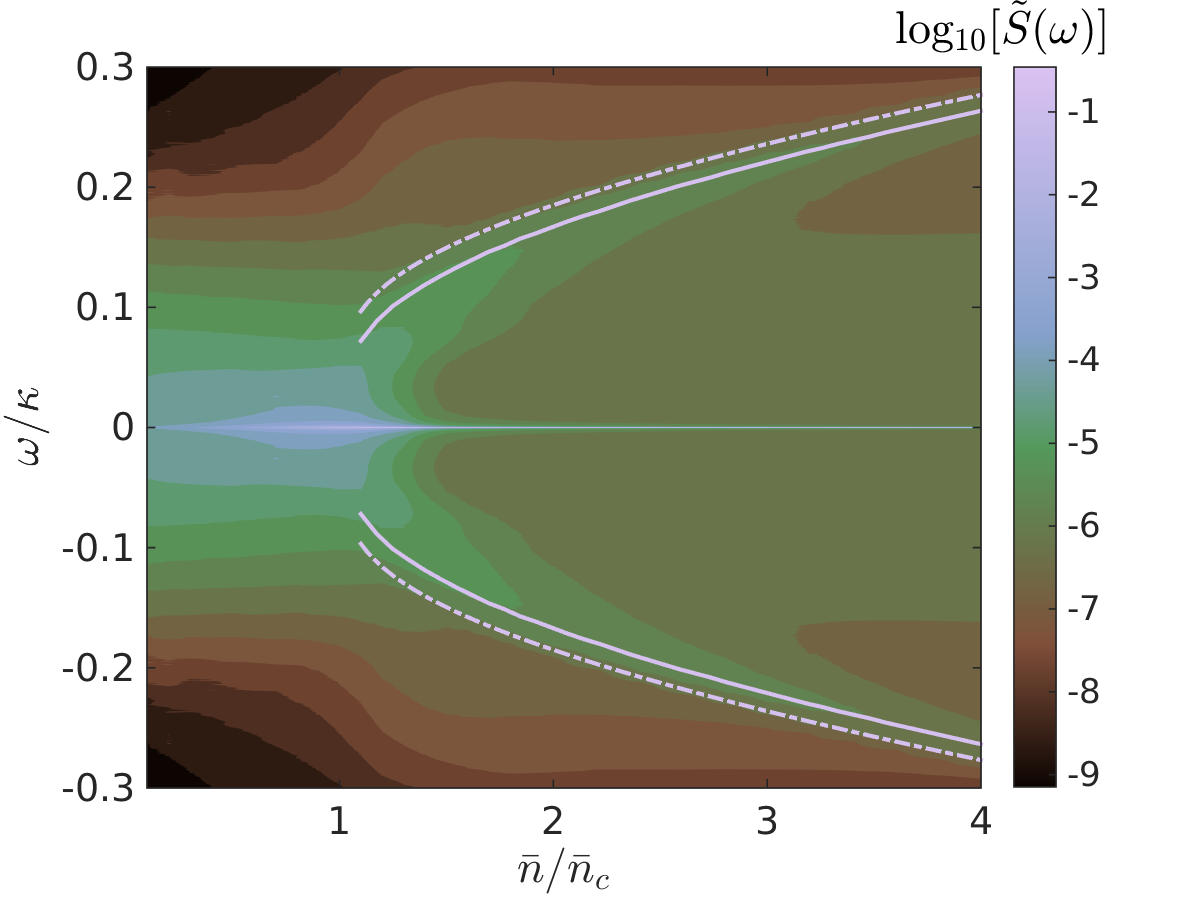} 
	\caption{ (Color online) \label{Wmax} Contour plot of the spectrum of the autocorrelation function $\tilde{S}(\omega)$ as a function of $\bar n$ and of the frequency (in units of $\kappa$) evaluated from the numerical data of  $\Theta (x_1,\ldots,x_N)$, Eq. \eqref{Theta:N}, by integrating the $N$-particle FPE, Eq. \eqref{FPE}, for 100 trajectories of $N=50$ atoms, $\Delta_c=-\kappa$, and evolution time $t_{\rm tot}=10^4\kappa^{-1}$, see Ref. \cite{Schuetz:2015}. The lines are analytical estimates of the spectrum maximum for $\bar n>\bar n_c$. The dashed line corresponds to the frequency of the corresponding harmonic oscillator in Eq. \eqref{omega0}. The solid line is at the frequency extracted by solving Eqs. \eqref{mean-field} for a mathematical pendulum and in good agreement with the peaks position of the numerically evaluated spectra. }
\end{figure}

The value of the intracavity photon number above threshold is found after observing that the exponent of function $\mathcal G(\alpha)$ has two minima that are given by the non vanishing solutions of the fixed-point equation \eqref{fixpointnew}, which we denote by $\Theta_{\mathrm{MF}}=\pm\bar{\Theta}$, with $\bar{\Theta}$ given in Eq. \eqref{Theta:0}. Therefore it holds
\begin{align*}
n_{\rm cav}=N\bar{n}\bar{\Theta}^2\approx 2N(\bar n-\bar n_c)\,,
\end{align*}
sufficiently close to the critical point. In particular, the mean number of photons increases linearly with $\bar n$.
We analyse now some properties of the first order correlation function of the intracavity field, $g^{(1)}(\tau)=\lim_{t\to\infty}{\rm Re}\langle E_{\rm cav}(t+\tau)E_{\rm cav}(t)\rangle_N$. This function has been extensively studied in Ref. \cite{Schuetz:2015} by numerically solving the $N$-particle FPE. Here, we will use the mean-field ansatz in order to better understand the two sidebands of its Fourier transform, at which it exhibits maxima above threshold. For this purpose we first notice that the correlation function is proportional to the autocorrelation function $C(\tau)$ of the magnetization by the relation $g^{(1)}(\tau)=N\bar n C(\tau)$, and
\begin{eqnarray}
C(\tau)&=&\lim\limits_{t\to\infty}\langle\Theta(t)\Theta(t+\tau)\rangle_N\,.
\end{eqnarray}
We want to derive $C(\tau)$ in mean-field and hence the mean value has now to be taken over the factorized distribution as in Eq. \eqref{factorization} with the stationary mean-field distribution given in Eq. \eqref{f:st}. We calculate $C(\tau)$ by solving the equations of the mathematical pendulum
\begin{align}
\dot{x}&=\frac{p}{m} \nonumber \\
\dot{p}&=2\hbar k \Delta_c\bar{n}\bar{\Theta}\sin(kx)\,, \label{mean-field}
\end{align}
with initial conditions $x(0)=x_0$ and $p(0)=p_0$. The value $\bar{\Theta}$ is here the positive stable solution of Eq. \eqref{fixpointnew}. In the limit of small oscillations, these equations describe harmonic motion at the frequency
\begin{align}
\omega_0=\sqrt{-4\omega_{r} \Delta_c\bar{n}\bar{\Theta}}\,. \label{omega0}
\end{align}
The mean frequency, however, is the result of the possible trajectories of the mathematical pendulum weighted by the probability density function $f_{\text{st}}(x_0,p_0)$. For $x_0\neq0$ and $p_0\neq0$ the oscillation period results to be larger than $2\pi/\omega_0$ and this prediction fits quite well the maximum found numerically by integrating the coupled equations of $N$ particles, as shown in Fig. \ref{Wmax}. %This further demonstrates the predictive power of the mean-field treatment for this system.

\subsection{Intensity-intensity correlations at zero-time delay}

\begin{figure}
	\includegraphics[width=1\linewidth]{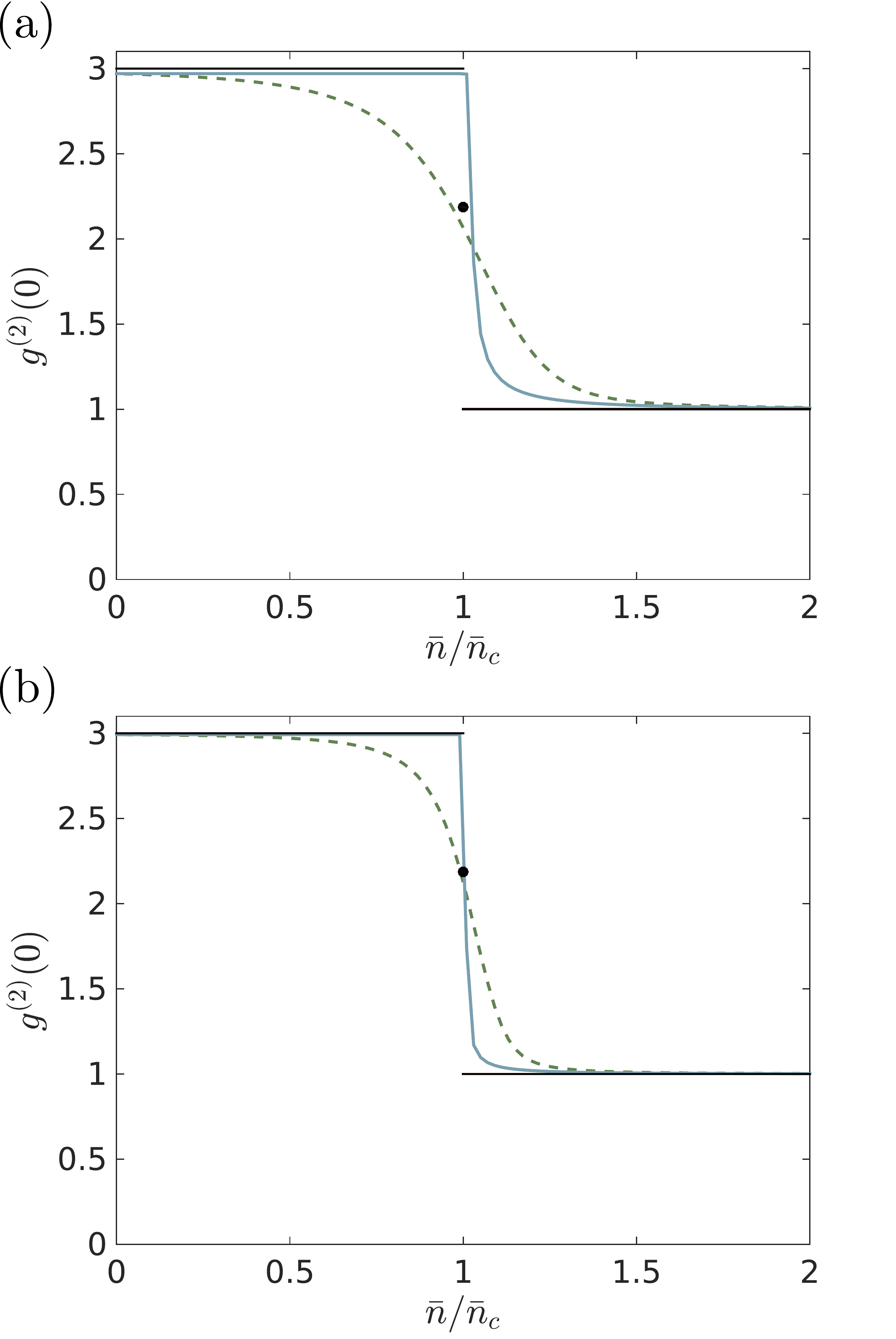}
	\caption{(Color online) \label{fig:g2}Intensity-intensity correlation function at zero time delay $g^{(2)}(0)$, Eq. \eqref{g:2}, as a function of $\bar n$ for (a) $N=50$ atoms and (b) $N=200$ atoms. The solid blue lines are the curves evaluated using in Eq. \eqref{g:2} the mean-field steady state \eqref{f:st}. The dashed lines are calculated for the corresponding full $N$-particle distribution given in \cite{Schuetz:2014}. The black solid lines are the values at the thermodynamic limit given at $\bar n<\bar n_c$ by Eq. \eqref{g:2:below} and at $\bar n>\bar n_c$ by Eq. \eqref{g:2:above}. The point at $\bar n=\bar n_c$ is at the value of Eq. \eqref{g:2:threshold}. The discrepancy between the mean-field curve and the full $N$-particle predictions decreases as $N\to \infty$, where they both converge to the value given by the thermodynamic limit. }
\end{figure}

The intensity-intensity correlation function at zero time delay,  $g^{(2)}(0)$, provides a direct measurement of the fourth moment of the magnetization when retardation effects are sufficiently small \cite{Schuetz:2015}:
\begin{equation}
\label{g:2}
g^{(2)}(0)=\left\langle\Theta^4\right\rangle_N/\langle \Theta^2\rangle_N^2\,.
\end{equation}
Above threshold $\langle \Theta^n\rangle_N=\bar{\Theta}^n+\mathcal{O}(1/N)$, with $\bar{\Theta}$ the solution of Eq. \eqref{fixpointnew}. Therefore, for $\bar n>\bar n_c$ we obtain
\begin{align}
\label{g:2:above}
g^{(2)}(0)_{\bar{n}>\bar{n}_c}=1\,,
\end{align}
which corresponds to coherent light and is valid at leading order, with an error that scales with $1/N$. In mean-field for the factorized distribution, Eq. \eqref{factorization}, we get 
\begin{align*}
\left\langle\Theta^2\right\rangle_N=\frac{1}{N}\mathcal{B}+\frac{N-1}{N}\bar{\Theta}^2
\end{align*} 
and 
\begin{align*}
\left\langle\Theta^4\right\rangle_N=&\frac{N(N-1)(N-2)(N-3)}{N^4}\bar{\Theta}^4\\
&+\frac{6N(N-1)(N-2)}{N^4}\bar{\Theta}^2\mathcal{B}+\frac{3N(N-1)}{N^4}\mathcal{B}^2\\
&+\frac{4N(N-1)}{N^4}\bar{\Theta}\langle\cos^3(x)\rangle+\frac{N}{N^4}\langle\cos^4(x)\rangle.
\end{align*}
Notice that above threshold for $\bar{\Theta}\neq0$ we can again write $\left\langle\Theta^4\right\rangle_N=\bar{\Theta}^4+\mathcal{O}(1/N)$.
Hence we get the same value for $g^{(2)}(0)=1$ (above threshold) in the thermodynamic limit $N\to\infty$.
Below threshold, in Appendix \ref{fluxexponent} we show that the expression takes the value
\begin{align}
\label{g:2:below}
g^{(2)}(0)_{\bar{n}<\bar{n}_c}=3\,,
\end{align}
which corresponds to super-Poissonian light. Corrections scale with $1/N$. The same holds for the calculation with the factorized ansatz. Below threshold we get 
\begin{align*}
\left\langle\Theta^2\right\rangle_N=\frac{1}{N}\mathcal{B}
\end{align*} 
and
\begin{align*}
\left\langle\Theta^4\right\rangle_N=\frac{3}{N^2}\mathcal{B}^2+\mathcal{O}\left(\frac{1}{N}\right)
\end{align*}
and therefore the same value of $g^{(2)}(0)=3$ (below threshold) as for the $N$-particle description.
Finally, at threshold we obtain
\begin{align}
\label{g:2:threshold}
g^{(2)}(0)_{\bar{n}=\bar{n}_c}\approx\frac{1}{4}\left(\frac{\Gamma\left[\frac{1}{4}\right]}{\Gamma\left[\frac{3}{4}\right]}\right)^2\,,
\end{align}
with corrections scaling with $1/\sqrt{N}$, thus giving a slower convergence than the one found for the values above and below threshold. We want to mention here that the mean-field description cannot reproduce the value in Eq. \eqref{g:2:threshold}. Figure \ref{fig:g2} displays the mean-field predictions for the $g^{(2)}(0)$ at the thermodynamic limit and as a function of $\bar n$. These curves are compared with the mean-field calculation at finite $N$ and with the corresponding one of the $N$-particle FPE. Even though the mean-field curve at finite $N$ is tendentially closer to the thermodynamic limit than the $N$-particle FPE prediction, they both converge to the values of Eqs. \eqref{g:2:above}, \eqref{g:2:threshold} \eqref{g:2:below}, depending on whether $\bar n<,=,>\bar n_c$, for $N\to \infty$.

\begin{widetext}
\subsection{Useful relations} \label{fluxexponent}
 In order to demonstrate Eq. \eqref{To:show} we first consider the relation 
  \begin{align*}
 \int_{-\infty}^{\infty}dy\exp\left(-\alpha N\left(y-\frac{1}{N}\sum_{i=1}^N\cos(kx_i)\right)^2\right)=\sqrt{\frac{\pi}{\alpha N}}\,,
 \end{align*}
 and cast it into the form
\begin{align*}
\int_{-\infty}^{\infty} dye^{-\alpha N y^2}\exp\left(2\alpha Ny\frac{1}{N}\sum_{i=1}^N\cos(kx_i)\right)
=\sqrt{\frac{\pi}{\alpha N}}\exp\left(\alpha N \Theta({\bf x})^2\right)\,.
\end{align*}
From these relations, it follows
\begin{align*}
\ln\left(\frac{1}{\lambda^N}\int d{\bf x}\exp\left(\alpha N \Theta({\bf x})^2\right)\right)=\frac{1}{2}\ln\left(\frac{ N}{\pi}\alpha\right)+\ln\left(\int_{-\infty}^{\infty} dy\exp\left[-N\left(\alpha y^2-\ln(I_0(2\alpha y))\right)\right]\right)\,.
\end{align*}
We use it for evaluating expression \eqref{n:cav} and obtain
\begin{eqnarray}
n_{\rm cav}=\bar{n}\frac{\partial}{\partial\alpha}\ln\left(\frac{1}{\lambda^N}\int d{\bf x}\exp\left(\alpha N \Theta({\bf x})^2\right)\right)
=\bar{n}\left(\frac{1}{2\alpha}+\frac{\frac{\partial}{\partial\alpha}\int_{-\infty}^{\infty} dy\exp\left[-N\left(\alpha y^2-\ln(I_0(2\alpha y))\right)\right]}{\int_{-\infty}^{\infty} dy\exp\left[-N\left(\alpha y^2-\ln(I_0(2\alpha y))\right)\right]}\right)\,,
\end{eqnarray}
that leads to Eq. \eqref{To:show} by using definition \eqref{G:alpha}.

In order to determine the intracavity photon number close to threshold, we expand the exponent of Eq. \eqref{G:alpha} about $y=0$ till fourth order:
\begin{align*}
\alpha y^2-\ln(I_0(2\alpha y))=\alpha(1-\alpha)y^2+\frac{\alpha^4}{4}y^4 + {\rm O}(y^6)\,.
\end{align*}
For $\bar n<\bar n_c$, the coefficient of the quadratic term is positive and we thus discard the fourth order term. Expression \eqref{To:show} takes the form
\begin{align*}
n_{\rm cav}\approx&\bar{n}\left(\frac{1}{2\alpha}+\frac{\frac{\partial}{\partial\alpha}\int_{-\infty}^{\infty} dy\exp\left[-N\alpha(1-\alpha)y^2\right]}{\int_{-\infty}^{\infty} dy\exp\left[-N\alpha(1-\alpha)y^2\right]}\right)\\
=&\bar{n}\left(\frac{1}{2\alpha}+\frac{2\alpha-1}{2\alpha(1-\alpha)}\right)=\frac{\bar{n}}{2(1-\alpha)}\,.
\end{align*}
Using the explicit value of $\alpha$, then 
\begin{equation}
n_{\rm cav}=\frac{\bar{n}\bar{n}_c}{2(\bar{n}_c-\bar{n})}\approx \frac{\bar{n}_c^2/2}{\bar{n}_c-\bar{n}}\,,
\end{equation}
which thus gives Eq. \eqref{n:cav:below}.

At the transition point $\bar{n}=\bar{n}_c$ the integral in Eq. \eqref{To:show} diverges in the limit $N\to\infty$. We determine its value for finite $N$, and keep the leading order. Moreover, since the coefficient of the quadratic term in the expansion in $y$ vanishes, we include the fourth order and evaluate the integral at $\alpha=1$, obtaining:
\begin{align*}
n_{\rm cav}\approx\bar{n}_c\left(\frac{1}{2}+\frac{\int_{-\infty}^{\infty} dy\left(Ny^2-Ny^4\right)\exp\left[-\frac{N}{4}y^4\right]}{\int_{-\infty}^{\infty} dy\exp\left[-\frac{N}{4}y^4\right]}\right)
\approx\bar{n}_c\frac{2\sqrt{N}\Gamma\left[\frac{3}{4}\right]}{\Gamma\left[\frac{1}{4}\right]}\,,
\end{align*}
which is the expression in Eq. \eqref{n:cav:at}.

To calculate $g^{(2)}(0)$ below and at threshold we notice that
\begin{align*}
N^2\left\langle\Theta^4\right\rangle_{N}-N^2\left\langle\Theta^2\right\rangle_{N}^2=\frac{\partial^2}{\partial\alpha^2}\ln\left(\frac{1}{\lambda^N}\int d{\bf x}\exp\left(\alpha N \Theta({\bf x})^2\right)\right)=N\frac{\partial}{\partial \alpha}\left\langle\Theta^2\right\rangle_{N}
\end{align*}
holds. Below threshold for $\alpha<1$ we calculated in leading order that
\begin{align*}
\frac{\partial}{\partial \alpha}\frac{1}{2(1-\alpha)}=\frac{1}{2(1-\alpha)^2}\,,
\end{align*}
which then delivers expression
\begin{align*}
g^{(2)}(0)_{\bar{n}<\bar{n}_c}=\frac{\frac{1}{2(1-\alpha)^2}+\frac{1}{4(1-\alpha)^2}}{\frac{1}{4(1-\alpha)^2}}=3\,,
\end{align*}
and thus Eq. \eqref{g:2:below}. In order to calculate the value at threshold we use
\begin{align}
N^2\langle\Theta^4\rangle_{N}-N^2\left\langle\Theta^2\right\rangle_{N}^2\approx&N-4N\left(\frac{\Gamma\left[\frac{3}{4}\right]}{\Gamma\left[\frac{1}{4}\right]}\right)^2\,,
\end{align}
which is valid in leading order and which gives Eq. \eqref{g:2:threshold}.
\end{widetext}

\end{document}